\documentclass{aa}  

\usepackage{graphicx}
\usepackage{natbib}
\usepackage{flexisym}
\usepackage{letltxmacro}
\LetLtxMacro\FlexiSymTextPrime\textprime
\DeclareTextCommandDefault{\textprime}{\FlexiSymTextPrime}

\usepackage{hyperref}

\usepackage{txfonts,textcomp}
\begin{document}

   \title{Ancient stellar populations in the outskirts of nearby grand-design spirals: Investigation of their star formation histories}

   \author{Cristina Maria Lofaro\inst{1,2}, Giulia Rodighiero\inst{1,3},  Andrea Enia\inst{4, 5}, Ariel Werle\inst{3}, Laura Bisigello\inst{1,3},  Paolo
   Cassata\inst{1,3}, Viviana Casasola\inst{5}, Alvio Renzini\inst{3}, Letizia Scaloni\inst{4,5}, Alessandro Bianchetti \inst{1,3}}
    
    \institute{Dipartimento di Fisica e Astronomia, Università di Padova, vicolo dell’Osservatorio 3, I-35122 Padova, Italy\\
              \email{clofaro@ia.forth.gr}
    \and
    Institute of Astrophysics, Foundation for Research and Technology–Hellas (FORTH), Heraklion, GR-70013, Greece
    \and
    INAF-Osservatorio Astronomico di Padova, vicolo dell'Osservatorio 5, I-35122, Padova, Italy
    \and
    University of Bologna, Department of Physics and Astronomy "Augusto Righi" (DIFA), Via Gobetti 93/2, I-40129, Bologna, Italy
    \and
    INAF-Osservatorio di Astrofisica e Scienza dello Spazio, Via Gobetti 93/3, I-40129, Bologna, Italy}

  \date{Received date /
    Accepted date }
   
   \titlerunning{Star formation history in nearby spirals}
   \authorrunning{Lofaro C. M. et al.}

   \abstract
    {The main sequence (MS) of star-forming galaxies (SFGs) is the tight relation between the galaxy stellar mass (M\textsubscript{$\star$}) and its star formation rate (SFR) and was observed up to z $\sim$ 6. The MS relation can be used as a reference for understanding the differences among galaxies, which are characterised by different rates of stellar production (starbursts, SFGs, and passive galaxies), and those inside a galaxy that is made up of different components (bulge, disk, and halo). To investigate peculiar features found in our sample galaxies in more depth, we focus here on their star formation history (SFH).}
    {The SFHs are a fundamental tool for revealing the galaxy path from the earlier stages of formation to the present time. The various phases of galaxy evolution are imprinted on the source spectrum globally and locally. Thus, we are able to interpret the dynamical origin of the spirals quantitatively and distinguish between in situ or ex situ formation processes.}
    {We performed a spectral energy distribution fitting procedure that accounted for the energetic balance between UV (observed) and far-IR (optically obscured) radiation on a sample of eight nearby face-on grand-design spiral galaxies from the DustPedia sample. This approach allowed us to study the spatially resolved MS of the sample and to recover the past SFH 
    by accounting for attenuation due to the presence of dust. By exploiting the BAGPIPES code, we constrained the SFHs for each galaxy with a delayed exponentially declining model to derive their mass-weighted age (t$\textsubscript{MW}$).} 
    {The spiral galaxies in our sample have similar radial t$\textsubscript{MW}$ trends overall. A central old region (t$\textsubscript{MW}$ up to $\sim7$Gyr, consistent with the presence of a bulge for various systems) is followed by younger regions in which the disks are still forming stars (t$\textsubscript{MW}\sim4$Gyr). At larger distances from the centre of the galaxies, t$\textsubscript{MW}$ increases mildly in general. Strikingly, in two galaxies (NGC4321 and NGC5194), 
    we found a steep increase in t$\textsubscript{MW}$ that reached levels similar to those of the bulge. These old stellar populations in the very galaxy outskirts, which are also detectable as "quenched rings" below the spatially resolved MS, is unexpected. We discuss their potential origin by considering the different gas phases (HI and H$_2$) of the source with the most prominent quenched ring, NGC4321, and argue for two main possibilities:  1) some environmental effect (e.g. starvation) could affect the outer edge of the galaxies or 2) the circumgalactic medium of sources outside of high-density clusters might have stopped to supply pristine gas to the galaxy (e.g. if its  specific angular moment is too high for being accreted).  
    }
    {}


   \maketitle
%

\section{Introduction}
Galaxies appear to build their stellar mass in a steady cold-gas accretion mode, as indicated by  the tight relation between the galaxy stellar mass (M$_*$) and its star formation rate (SFR): the main sequence (MS) of star-forming galaxies (SFGs), observed up to $z\sim6$ with a fairly constant scatter of $\sim$0.3 dex \citep[see, e.g.][]{Noeske,Daddi,Rodighiero,Speagle,Whitaker,Wuyts,Schreiber,Renzini,Popesso}.
Galaxies seem to fluctuate around the MS relation as a consequence of stochastic variations in their gas accretion rate and ensuing  SFR, with occasional more dramatic events such as disk instabilities \citep{Tacchella16}. One outstanding issue remains: to  understand how the star formation activity is eventually quenched and leads to the bimodality between star-forming and passive galaxies 
\citep[e.g.][]{Rodighiero1,Peng,Saintonage}.
The tight relation between stellar mass surface density ($\Sigma_*$) and SFR
surface density ($\Sigma\textsubscript{SFR}$ ) found in HII regions of nearby galaxies further suggests that the global MS relation originates in local processes that set the conversion of gas into stars \citep{Rosales-Ortega,SA}. 
Following this, several works have exploited the advent of large integral field spectroscopic surveys to analyse the spatially resolved MS relation in low-redshift galaxies using the $H\alpha$ flux as SFR tracer \citep[e.g.][]{Cano-Diaz,Hsieh,Abdurro’uf,Medling}.

\citet{Enia} studied the main biases affecting most previous works in which the SFRs were derived from emission lines (i.e. $H\alpha$) or UV-to-optical tracers \citep[e.g.]{SA}, and which therefore likely underestimated the SFR of more dust-extincted regions.  To achieve this, they made use of a more complete photometric coverage that extended from the far-UV to the far-IR (i.e. from GALEX to Herschel) to provide a reliable measure of the SFR in galaxies by directly accounting for both the observed and the dust-obscured components. This can be performed by a spectral energy distribution (SED) fitting procedure that accounts for the energetic balance between dust-absorbed and re-emitted radiation, hence offering a more complete account of the SFRs. The sample presented in \citet{Enia}  is restricted to grand-design spirals with a low inclination, large spatial extension, and regular spiral arms structures. This analysis was extended to ten thousands of physical cells on typical scales of $\sim$0.5 kpc and  over very different internal galactic regions (bulges, spiral arms, inter-arms regions, and
outskirts). This set of galaxies offers a local reference to explore  the evolutionary processes regulating  star
formation in rotationally supported systems all the way to their high-redshift counterpart \citep{Law,Schreiber1,Glazebrook,Wuyts13,Simons,Schreiber2,Ubler}.
However, individual galaxies show specific variations around the $\Sigma\textsubscript{SFR}$-$\Sigma_*$ relation, indicating that a comprehensive analysis of a larger sample is required.
Before we attempt this extension, we take one step farther here than \citet{Enia} by estimating the star formation histories (SFH) \citep[see][for a review]{Walcher} of individual galactic cells on the same 0.5 kpc scale.

We therefore undertake a comprehensive analysis of the spatially resolved MS of a sample of nearby grand-design spiral galaxies in which, while following a global scaling relation on average  \citep{Enia,Ellison,Casasola22}, some galaxies
can reveal peculiar features that are not observed in others. 
In particular, in some physical regions within our galaxies, the observed SFR significantly exceeds that predicted by the main relation at fixed stellar mass (i.e. starbursting regions). On the other hand, we observe some areas that are characterised by a suppression of the SFR at a given stellar mass, which we consider as quenched regions (lying up to more than 1 dex below the expectation of the star-forming main sequence).
We focus on these latter regions with the goal of using spatially resolved SFHs to understand the physical origin of the quenching mechanisms in star-forming galaxies.
We limited our analysis to eight sources as a pilot program
to test the method, which we plan to extend in the future to a much wider galaxy sample. 

The paper is structured as follows. In Section 2 we describe the dataset. In Section 3 we present our SED fitting
method (e.g. libraries and image processing). In Section 4
we present our results for the spatially resolved MS. We assume a $\Lambda$CDM cosmology \citet[][]{Planck} and \citet[][]{Chabrier} initial mass function (IMF) throughout. 

\section{DUSTPEDIA sample}

   
%
\subsection{DustPedia archive}
DustPedia is a research project to characterise the dust in the local Universe. Posed as the legacy of the Hershel Space Observatory, it contains the imagery and photometry study of 875 nearby galaxies, spanning over five orders of magnitude in wavelength, from the ultraviolet to the microwave, and  was made publicly available to the scientific community through the DustPedia database \citep{Clark} \footnote{The DustPedia website is available at http://dustpedia.astro.noa.gr}.
Subsequently, the considerable data improvement of the past years and exploration of a wide spectral range (from the far-infrared to the sub-millimetre) enabled a relevant development in the cosmic dust investigation through the Herschel, Planck, Spitzer, the James Clerk Maxwell Telescope (JCMT), and the Atacama Large Millimetre/sub-millimetre Array (ALMA) observatories. These observatories were well suited for the study of nearby galaxies through their rapid mapping abilities, which enabled them to observe a sizeable portion of the galaxies in the local Universe with mixed sensitivity and resolution, and with a broad wavelength coverage.\\
The Dustpedia sample consists of galaxies observed within a distance of 41 Mpc assuming $H\textsubscript{0}$=73.24 km/s/Mpc and $D\textsubscript{25}$ > 1$\textsuperscript{$\prime$}.$ $D\textsubscript{25}$ is the major axis isophote at which the optical surface brightness falls beneath $25 mag/arcsec^2$.

\subsection{Multiwavelength data}
The ancillary  data consist of observations from the following
facilities (for further details, see Tab \ref{tab:1}): \\ - the \textit{GALaxy Evolution eXplorer}, GALEX \citep{Morrissey}. The ultraviolet part of the electromagnetic spectrum is sampled by GALEX. Near-UV (NUV) and far-UV (FUV) data sample the light from newborn massive stars that traces the unobscured star formation activity of
galaxies. \\-  the \textit{Sloan Digital Sky Survey} \citep[SDSS;][]{York,Eisenstein}. The SDSS provides ultraviolet, optical, and near-infrared imaging of the 35 $\%$ of the sky and samples the young stellar content. \\ 
- \textit{the 2 Micron All-Sky Survey}, 2MASS \citep{Skrutskie}, \\ 
- the \textit{Wide-field Infrared Survey Explorer}, WISE \citep{Wright}, and\\
- the \textit{Spitzer Space Telescope} \citep{Werner}. 
The NIR and MIR observations from the 2MASS, WISE, and Spitzer surveys trace the old stellar component, the stellar mass distribution, and the carbonaceous-to-silicate materials in the dust. \\ 
- {\it Herschel Space Observatory} \citep{Pilbratt} and {\it Planck}  \citep{Planck}. {\it Herschel} and {\it Planck}  cover the spectral range  from the far-infrared up to the sub-millimeter, which allows us to probe the reprocessed emission from dust and to thus constrain the dust-obscured star formation processes. 

\subsection{HI and CO observations}
To further support the interpretation of our results, we took advantage of atomic and molecular gas observations, in particular, for galaxy NGC4321 (see Section \ref{sample}).
HI data are available from the public VIVA survey \citep[VLA IMAGING OF VIRGO SPIRALS IN ATOMIC GAS,][]{Chung}. These observations were carried out with the Very Large Array (VLA) and are characterised by an angular resolution of $\sim15"$.
To compute the HI surface brightness $\Sigma_{HI}$, we followed the approach detailed in \cite{Morselli}.  We convolved the 21cm natural-weighted intensity maps, given in Jy beam$^{-1}$m s$^{-1}$, to the resolution of the worst of the 23 photometric bands used in the SED fitting  \citep[the one of SPIRE350, 24 arcsec, see][]{Enia} using a Gaussian kernel. 
Consistently with \cite{Morselli}, we computed the sensitivity limit as the rms of the $\Sigma_{HI}$ maps over an aperture corresponding to a diameter of 500pc, which we measured as log$\Sigma_{H_I,lim}=$0.3 M$_\sun$ pc$^{-2}$.

The molecular gas surface density, $\Sigma_{H_2}$, was computed using the $^{12}$CO(2-1) intensity maps from the HERACLES survey \citep[The HERA CO-Line Extragalactic Survey,][]{Leroy}. These observations were made with the IRAM 30m telescope and have an angular resolution of 11 arcsec. As for $\Sigma_{HI}$, we convolved the images using a Gaussian kernel to the resolution of SPIRE350.
We refer to \cite{Morselli} for a detailed description of the assumptions that we applied to convert CO fluxes into $\Sigma_{H_2}$ measurements.
We computed the sensitivity limit as the rms of the $\Sigma_{H_2}$ maps over an aperture corresponding to a diameter of 500pc, which we measured as log$\Sigma_{H_2,lim}$= 0.4 M$_\sun$ pc$^{-2}$.
\begin{table}[htbp]
  \centering
  \begin{tabular}{p{1cm}*{3}{c}}
    \noalign{\smallskip}
    Facility & Effective wavelength & Pixel width \\
    \empty & \empty & ($''$) \\
    \hline
    \noalign{\smallskip}
    GALEX & 153 nm & 3.2 \\
    \noalign{\smallskip}
    GALEX & 227 nm & 3.2 \\  
    \noalign{\smallskip}
    SDSS & 353 nm & 0.45\\
    \noalign{\smallskip}
    SDSS & 475 nm & 0.45\\
    \noalign{\smallskip}
    SDSS & 622 nm & 0.45\\
    \noalign{\smallskip}
    SDSS & 763 nm & 0.45\\
    \noalign{\smallskip}
    SDSS & 905 nm & 0.45\\
    \noalign{\smallskip}
    2MASS & 1.24 $\mu$m & 1 \\
    \noalign{\smallskip}
    2MASS & 1.66 $\mu$m & 1 \\
    \noalign{\smallskip}
    2MASS & 2.16 $\mu$m & 1 \\
    \noalign{\smallskip}
    WISE & 3.4 $\mu$m & 1.375 \\  
    \noalign{\smallskip}
    WISE & 4.6 $\mu$m & 1.375 \\  
    \noalign{\smallskip}
    WISE & 12 $\mu$m & 1.375 \\  
    \noalign{\smallskip}
    WISE & 22 $\mu$m & 1.375 \\  
    \noalign{\smallskip}
    \textit{Spitzer} & 3.6 $\mu$m & 0.75\\
    \noalign{\smallskip}
    \textit{Spitzer} & 4.5 $\mu$m & 0.75\\
    \noalign{\smallskip}
    \textit{Spitzer} & 5.8 $\mu$m & 0.6\\
    \noalign{\smallskip}
    \textit{Spitzer} & 8.0 $\mu$m & 0.6\\
    \noalign{\smallskip}
    PACS & 70 $\mu$m & 2\\  
    \noalign{\smallskip}
    PACS & 100 $\mu$m & 3\\
    \noalign{\smallskip}
    PACS & 160 $\mu$m & 4\\
    \noalign{\smallskip}
    SPIRE & 250 $\mu$m & 6\\  
    \noalign{\smallskip}
  \end{tabular} 
  \caption{Name, effective wavelength, and pixel width of the DustPedia facilities.} 
  \label{tab:1}
\end{table}

\subsection{Galaxy sample}
\label{sample}
From the many galaxies in the DustPedia archive, we chose a sample of eight grand-design spiral galaxies that were previously selected in \citet{Enia}: NGC0628,
NGC3184, NGC3938, NGC4254, NGC4321, NGC4535,
NGC5194, and NGC5457. The sample includes different morphologies and environments: isolated galaxies (NGC3184, NGC0628, and NGC3938), cluster galaxies (NGC4321, NGC4254, NGC4535, and NGC5457), and galaxies involved in merger processes (NGC4254 and NGC5194). 
Further details of each galaxy are shown in Table \ref{tab:2}. 
  
\begin{table*}[htbp]
  \centering
  \begin{tabular}{p{4cm}*{10}{c}}
    \noalign{\smallskip}
    Galaxy name & RA & DEC & D & i & r\textsubscript{25} & logM\textsubscript{$\star$} & SFR & RC3 Type & Cell size \\
    \empty & [deg] & [deg] & [Mpc] & [$\circ$] & [kpc] & [M\textsubscript{$\odot$}] & [M\textsubscript{$\odot$} / yr] & \empty & 8$''$[kpc] \\
    \hline
    \noalign{\smallskip}
    NGC0628 &  24.1740 & 15.7833 & 10.14 & 19.8 &  14.74 &  10.41$\pm$0.15 &  1.90$\pm$0.41 &  Sc &  0.39 \\  
    \noalign{\smallskip}
    NGC3184 & 154.5708 & 41.4244 & 11.64 & 14.4 & 12.55 & 10.14$\pm$0.10 & 0.98$\pm$0.10 & SABc & 0.45 \\
    \noalign{\smallskip}
    NGC3938 &  178.2057 & 44.12088 &  19.41 &   14.1 &   10.04 &    10.16$\pm$0.20 &    2.19$\pm$0.19 &  Sc &    0.75 \\  
    \noalign{\smallskip}
    NGC4254(M99) &   184.7065 &  14.4164 &    12.88 &    20.1 &    9.40 &    10.02$\pm$0.18 &      2.44$\pm$0.23 &   Sc &   0.50 \\  
    \noalign{\smallskip}
    NGC4321 & 185.7282 & 15.8219 & 15.92 & 23.4 &  14.30 & 10.74$\pm$0.15 & 3.27$\pm$0.37 & SABb & 0.62 \\  
    \noalign{\smallskip}
    NGC4535 &  188.5845 & 8.1978 & 14.93 &  23.8 &  17.62 &   10.19$\pm$0.19 &   1.30$\pm$0.08 &  Sc &   0.58 \\  
    \noalign{\smallskip}
    NGC5194 (M51) & 202.4695 & 47.1952 & 8.59 & 32.6 & 17.23 & 10.70$\pm$0.20 & 4.08$\pm$0.26 & Sbc & 0.33\\
    \noalign{\smallskip}
    NGC5457(M101) &  210.8025 &  54.3491 &   7.11 &    16.1 &   24.81 &    10.38$\pm$0.13 &     2.48$\pm$0.15 &  SABc &    0.28 \\  
    \noalign{\smallskip}
  \end{tabular} 
  \\
  \caption{DustPedia galaxy sub-sample. The galaxy name, coordinates in J2000 system reference, distances D (in Mpc), inclinations, r\textsubscript{25} sizes, and morphological classifications shared in the DustPedia archive come from the HyperLEDA database \citep{Makarov}. The value of M\textsubscript{$\star$} was obtained by fitting the DustPedia photometry with the SED-fitting code MAGPHYS, and the SFRs values were computed in \citet{Enia}.}
  \label{tab:2}
\end{table*}

In addition to the morphological classification and the environment, the selection of the eight galaxies was also based on inclination, distance, and photometry. In order to limit corrections for dust and/or disk inclination, the galaxy inclination \textit{i} was limited to less than 40°, so that they can be said to be nearly face-on galaxies. Furthermore, a cutoff in distance was applied (around 1000 km/s, corresponding to approximately 22 Mpc) because for galaxies beyond this distance, the galaxy is scarcely resolved in the sub-millimeter. Finally, to robustly perform an SED fitting study, each galaxy of the sample was observed in at least 20 bands, based upon which, it is possible to estimate their physical parameters.

\subsubsection{Image processing, data reduction, and flux evaluation}
For each galaxy of the sample, the DustPedia archive contains multi-wavelength photometric observations (flux and error maps). For each map, the background estimation and subtraction and the PSF degradation were performed. The background estimation and subtraction procedure follows the indications given in \cite{Clark} (see \cite{Enia} for further details); therefore, the background-subtracted maps were degraded to the SPIRE350 PSF ($8 \textprime\textprime $). To do this, the maps were convolved using the kernels provided by \cite{Aniano} (see \cite{Enia} for further details). We analysed the galaxies through an SED-fitting procedure by considering $8 \textprime\textprime $ x $8 \textprime\textprime $ cells. \\
Finally, in order to measure the flux in each band inside the apertures, the \textit{photutils v0.6 Python} package \cite{larry_bradley_2023_1035865} was used. On the other hand, the error can be estimated, when available, from the Dustpedia database. When this is not possible, the signal-to-noise ratio (S/N) of the DustPedia photometry in that particular band can help to compute the error in each cell. As an assumption, the S/N threshold was equal to 3: pixels lower than this value were ignored. Moreover, for a pixel for which the flux measurement could not be performed in more than ten bands, the pixel was rejected (see \cite{Enia} for further details).  

\section{Results and discussion}
\subsection{Spatially resolved star formation histories}
\label{SFH}
The motivation of our work is understanding the origin of the significant deviations from the average MS that were observed in the
sample presented in \citet{Enia}.
For the purpose of our studies, a further investigation of the
SFH of the galaxies examined is therefore highly recommended. This was allowed through the BAGPIPES SED-fitting code \citep{Carnall}. 
An example of an SED plot for three representative apertures (at a distance of $0$, $0.5 R_{25}$, and $R_{25}$) of the NGC4321 galaxy is provided in \ref{galaxy-outputs}.
We note that in the original paper \citep{Enia}, the physical parameters of the sources were derived with MAGPHYS \citep{daCunha}. However, this code does not allow us to easily recover the adopted SFH for each object. We therefore relied for this work on the more flexible BAGPIPES fitting procedure. However, we found that the two codes provide almost consistent results in terms of the derived stellar masses and SFR (see \ref{Comparison} for quantitative details).

In the following, we refer to stellar masses  derived from BAGPIPES, while SFRs (as in \citet{Enia} and \citet{Morselli}) are obtained as SFR = SFR$_{UV}$ + SFR$_{IR}$. We note that SFR$_{UV}/(M\textsubscript{$\odot$}/yr) = 0.88 \times 10\textsuperscript{-28}\, \textit{L}_\nu/\textit{L}_{\odot}$
with \textit{L}\textsubscript{$\nu$}, in \textit{erg/s/Hz}, as the luminosity per unit frequency evaluated at 150 $n$m \citep[from][]{Bell} and SFR$_{IR} = 2.64 \times 10\textsuperscript{-44}\, \textit{L}_{IR}$ with \textit{L}\textsubscript{IR}, in \textit{erg/s}, as the luminosity evaluated from the SED fit between 8$\mu$m and 1000$\mu$m \citep[][]{Kennicutt}.

\subsubsection{Delayed exponentially declining model}
We adopted a conservative approach by exploring the sample through a delayed exponentially declining SFH model, 
\begin{equation}
 SFR(\textit{t}) \propto
\left\{
\begin{array}{rl}
(\textit{t}-\textit{T}\textsubscript{0})\: exp \:(- \frac{\textit{t}-\textit{T}\textsubscript{0}}{\tau}) &  \textit{t}>\textit{T}\textsubscript{0}\\
0 &  \textit{t}<\textit{T}\textsubscript{0}

\end{array}  
\right.    
\end{equation},
where t is the time since the Big Bang, $T_0$ is the time when star formation starts, and $\tau$ is the timescale of the exponentially declining star formation. 

We adopted the following specifications for the model parameters: $T_0$
varying from 
0.1 to 9 Gyr \citep[since the Big Bang, to avoid stellar ages younger than $\sim$5Gyr, that would be inconsistent with the oldest stellar populations observed in local spiral galaxies, e.g.][and references therein]{Peterken20}, 
$\tau$
ranging from 1 to 10 Gyr (again since the Big Bang), the total mass formed over the whole history (in $log_{10}(M_{\star}/M_{\odot}$)) spanning from 1 to 15, and metallicities varying between 0 and 2.5 $Z_{\odot}$ \citep[for further details, see][]{Carnall}. 
The shape of the attenuation curve is provided by \cite{Calzetti}, with a V-band attenuation (A$_V$) from 0 to 2 mag. The redshift was finally provided by the DustPedia database.

We divided each galaxy in apertures with diameters of 8" \citep[see][for more details]{Enia} , and we performed the SED-fitting procedure in each of them. 
As a first step, we retrieved the stellar mass surface density and the SFR surface density, and we calculated the distance from the main sequence for each aperture. 

Fig.\ref{fig:NGC4321} shows (upper panels) as an example the $\Sigma\textsubscript{$\star$}$ and $\Sigma\textsubscript{SFR}$ of  NGC4321, that is, the stellar mass density and SFR density.  
The two were obtained by dividing the M\textsubscript{$\star$} and SFR quantities with the area given by the square of the side aperture (8$''$-side aperture). 
As previously mentioned, the stellar mass surface density was computed as an output of
the SED fitting, while the SFRs used in the SFR surface density were still computed using the $ SFR = SFR_{UV} + SFR_{IR}$. 


The stellar mass and SFR distributions clearly both have a marked pick in the centre and along the spiral arms. This shows that both the M\textsubscript{$\star$} and the SFR trace the spiral pattern of the galaxies. 
In the same figure, we also show again for NGC4321 the distance from the MS of each cell (bottom left panel) and the spatially resolved MS, highlighting the distance of each cell from the galaxy centre (bottom right panel). 
It is worth pointing out that the galaxy has a very
remarkable quenched ring: The portions of NGC4321 at the lowest stellar mass densities (i.e. at large galactocentric distances) clearly lie below the MS, which is an indication of suppressed star formation.

The spatially resolved MS plot (Fig.\ref{fig:NGC4321}, bottom right panel), NGC4321 presents a skewed distribution in the stellar mass-SFR relation: At log($\Sigma_*$)$>7.0$ M${_\odot}$kpc$^{-2}$, the data follow the average MS scaling relation,  whereas at lower stellar mass densities, the data sit well below the expected values considering the average of all points across the sample. At fixed stellar mass, the observed SFR is suppressed, indicating quenched galactic regions. 

The original sample of eight face-on spirals  (see \ref{galaxy-outputs} for the galaxy-by-galaxy results) contains other galaxies with a potential indication of a lower specific SFR toward the outskirts (e.g. NGC4535 and NGC5194). This corresponds to  $\sim40\%$ of the total (including NGC4321). Even within the limited statistic available to us, it is clear that this peculiar feature is not rare.
It is thus crucial to characterise the physical properties of the stellar populations in these under-dense and passive external regions of the galaxies. We therefore explored the information that can be drawn from a systematic study of the SFHs, taking advantage of the performances of the BAGPIPES fitting procedure.

 \begin{figure*}[htpb]
        \centering
        \begin{minipage}{1\columnwidth}
                \centering
                \includegraphics[width=\textwidth]{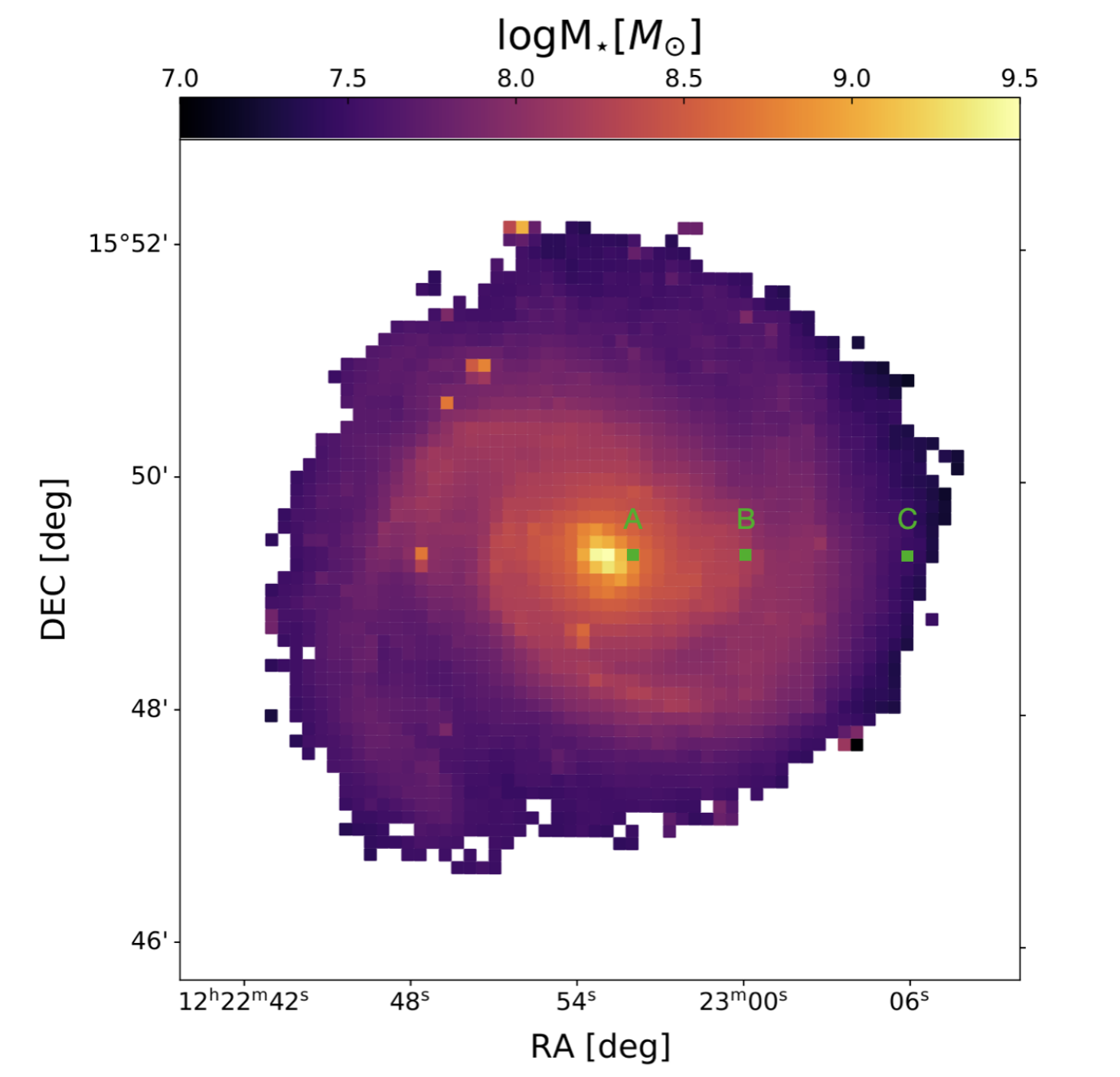}
                
                \label{label1}
        \end{minipage}%
        \begin{minipage}{1\columnwidth}
                \centering
                \includegraphics[width=\textwidth]{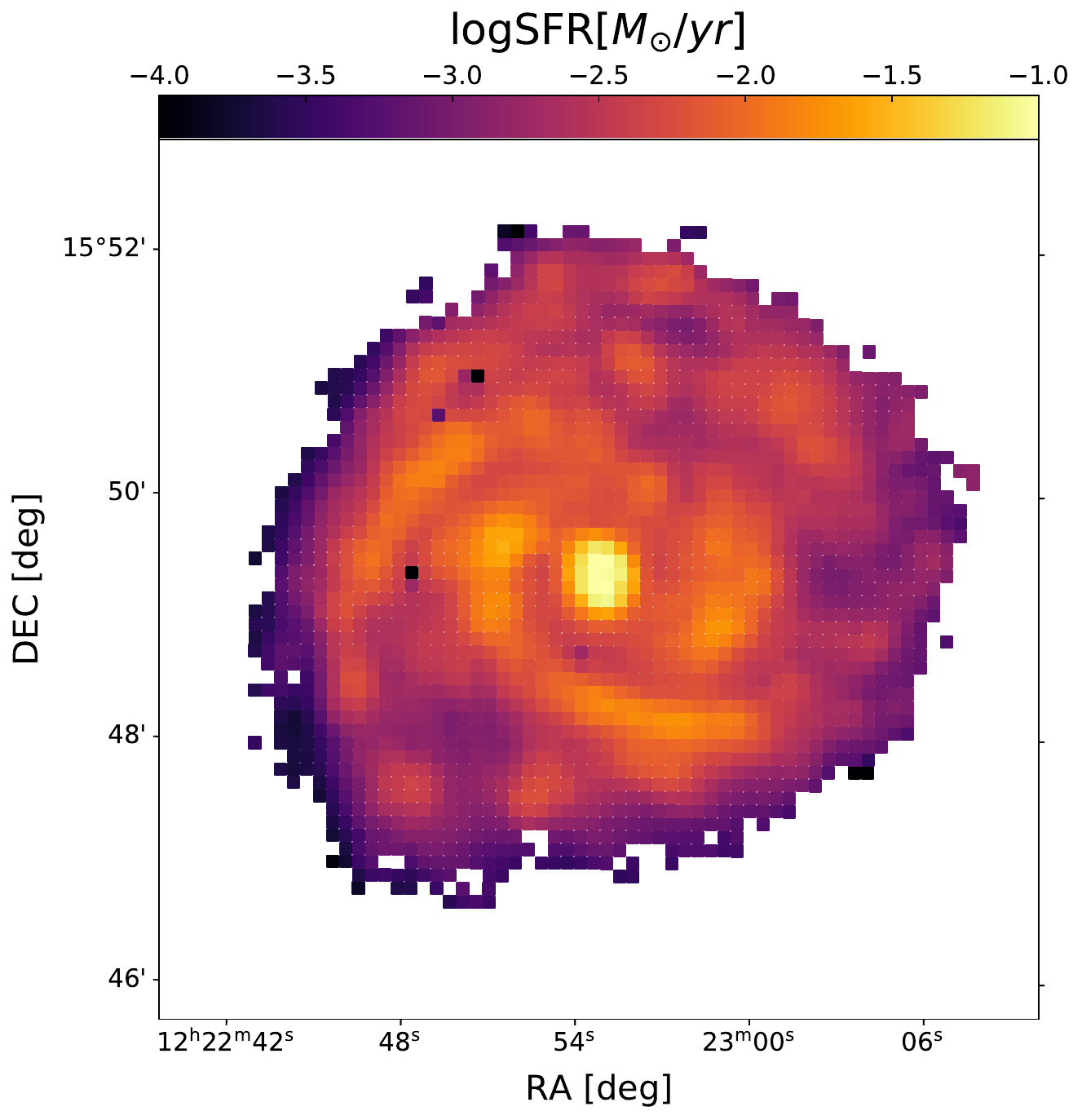}
                
                \label{label2}
        \end{minipage}
 \end{figure*}

 \begin{figure*}[htpb]
        \centering
        \begin{minipage}{1\columnwidth}
                \centering
                \includegraphics[width=\textwidth]{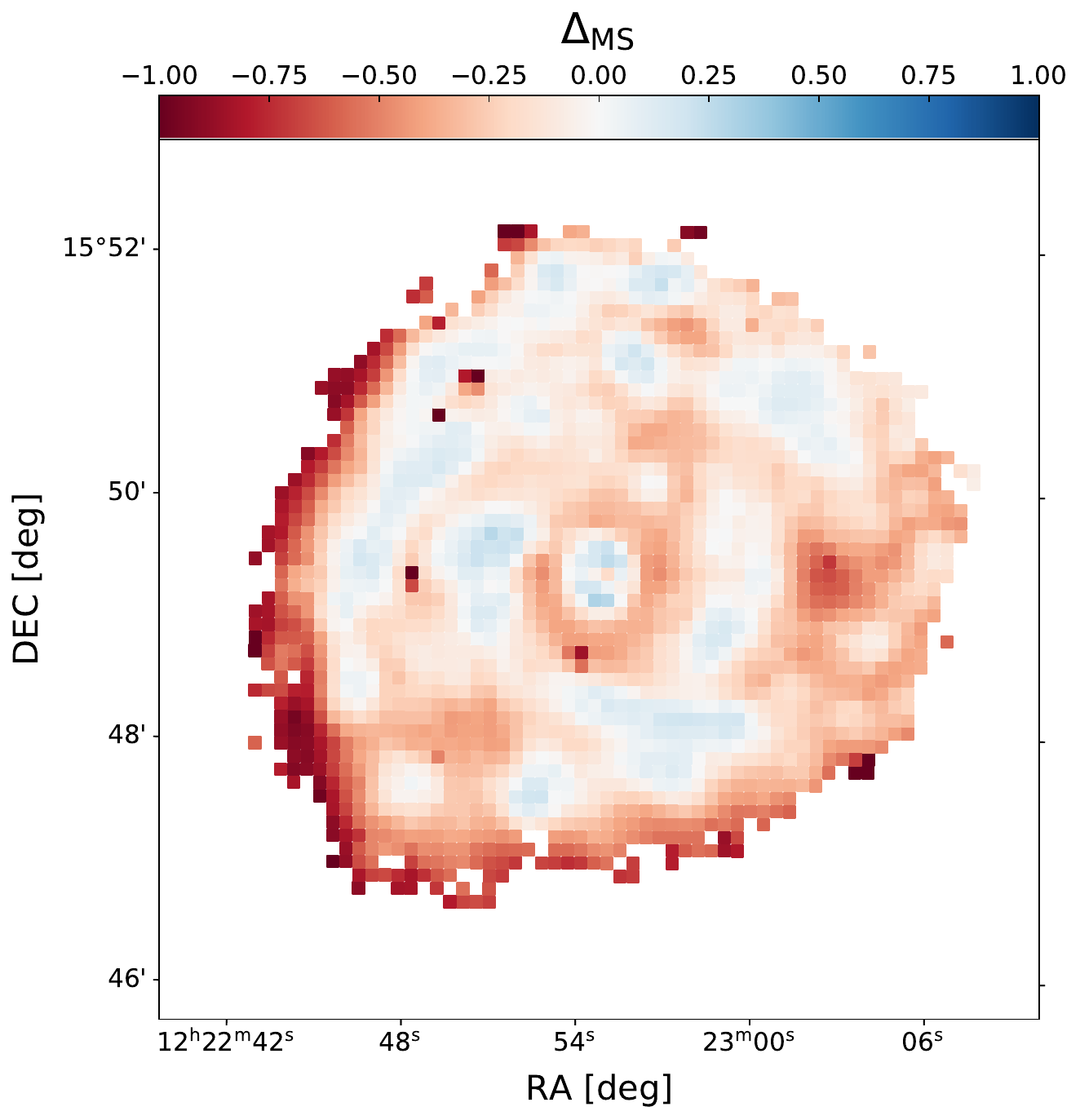}
                
                \label{label3}
        \end{minipage}%
        \begin{minipage}{1\columnwidth}
                \centering
                \includegraphics[width=\textwidth]{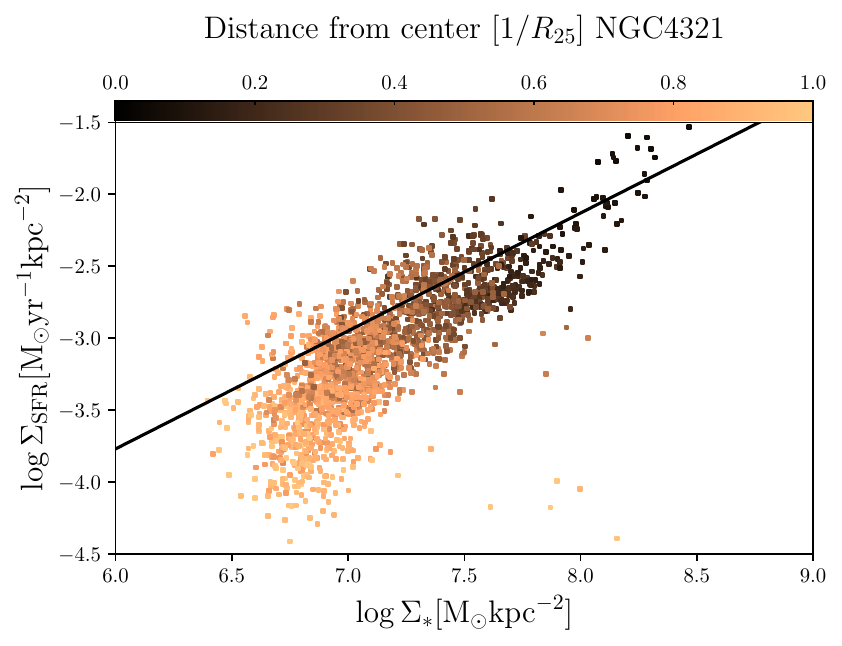}
        
                \label{label4}
        \end{minipage}
        \caption{Summary plot for NGC4321. The panels are organised as follows: stellar mass surface density (\textit{upper left}), SFR surface density (\textit{upper right}), distance from the MS (\textit{bottom left}), and spatially resolved MS (\textit{bottom right}). The solid black line shows the spatially resolved main sequence by \citet{Enia}, and each point is colour-coded by the cell distance from the galaxy centre. In the first panel, three apertures (A, B, andC) are highlighted. The corresponding SED plot for each is reported in Appendix A. }
        \label{fig:NGC4321}
 \end{figure*}


\subsection{Mass-weighted ages for the full sample}
\label{mass-weighted age}
\begin{figure*}
   \sidecaption
   \includegraphics[trim=0 220 40 220,clip,width=1.4\columnwidth]{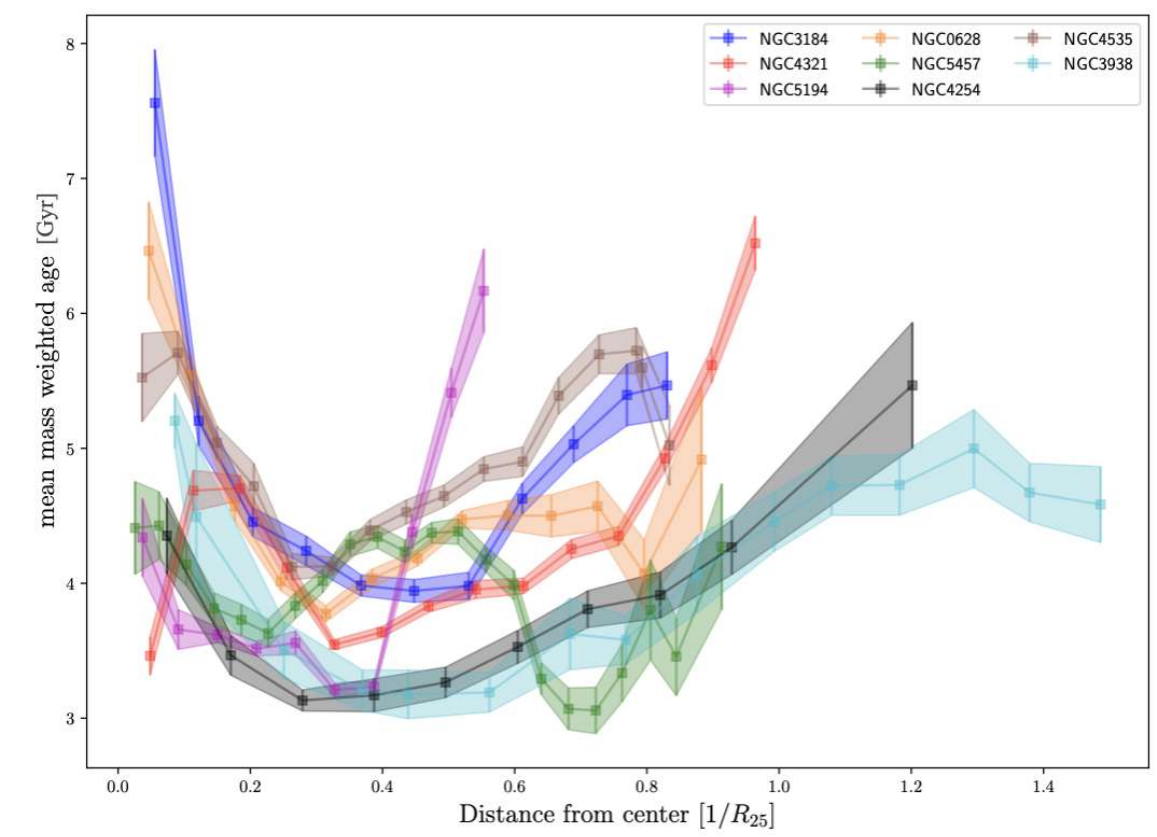}
      \caption{Average radial distribution of the mass weighted age of the eight galaxies of our sample. The distance from the center is normalised to their R\textsubscript{25} radius. As regards the uncertainties computation, see Sec. \ref{mass-weighted age}}
         \label{fig:mwa_all}
   \end{figure*}

In this section, we discuss the implications of the SED fitting results from the BAGPIPES code. By selection, the galaxies in the sample \citep{Enia} represent the final product of secular processes that convert the gas into stars. Spiral galaxies are built across the Hubble time: They form stars at a pacific rate (excluding major merger events), driven by the density wave turbulence crossing the interstellar medium within the gaseous disks.  The regulation of this smooth star formation mode is at the origin of the observed main sequence (MS) of star-forming galaxies and holds at different spatial scales (from $\sim$0.5 kpc to the whole galaxy). \\
Here, we specifically focused on the apertures of each galaxy underlying the MS in their peripheral regions (see Appendix B for the plots showing the distance from the MS for each galaxy). We report the results of the mass-weighted age (t$\textsubscript{MW}$) as an output from BAGPIPES.
t$\textsubscript{MW}$ gives an indication of the epoch at which the stellar masses of galaxies were assembled, and it was computed by weighting the mass of stars at the time of formation,
\begin{equation}
    t\textsubscript{MW} = \frac{\int_{0}^{t\textsubscript{obs}}t SFR(t) \textit{dt}}{\int_{0}^{t\textsubscript{obs}}SFR(t) \textit{dt}},
\end{equation}
where \textit{SFR(t)} is the SFH, and $t\textsubscript{obs} = t(z\textsubscript{obs})$. 
Consequently, for a visual inspection and to derive more quantitative conclusions, we derived the radial profiles of the mass-weighted age for each galaxy (see \ref{galaxy-outputs}). 
In particular, we show in Fig. \ref{fig:mwa_all} the radial profiles obtained by averaging all pixels at the same distance from the corresponding galaxy centre for each galaxy separately. The galactocentric distances are normalised to the R\textsubscript{25} radius for a more meaningful comparison of sources with different physical sizes and at different distances (this approach to radial profiles of different properties of DustPedia galaxies was also used in \citet{Casasola17}. 
The statistical errors on $t_{MW}$ were computed as the standard deviation on the mean value in each distance bin.
 To also account for the effect of the model uncertainties on $t_{MW}$,  we used the information included in the posterior distributions obtained by the BAGPIPES run. For each cell, we derived the difference between the 84th and 16th percentile of the $t_{MW}$ distribution, 
 and assumed half of it as the error on each $t_{MW}$ ($err_{t_{MW}}$=($t_{MW}$(84th)-$t_{MW}$(16th))/2). 
We assumed the median of the distributions of $err_{t_{MW}}$ for all the cells in each radial bin as a systematic error that was added in quadrature to the statistical error on the average $t_{MW}$ in each radial bin. 

We point out that the mass-weighted age ($t_{MW}$) was not analysed in absolute terms because it is well known that an underestimation of mass-weighted ages can be observed in SED-fitting analyses that adopt both parametric and non-parametric SFHs \citep[][among others]{Carnall}. 
We rather aim at detecting relative variations of $t_{MW}$ within each galaxy and at deriving average trends as a function of the distance from the centre.

The radial distribution of $t_{MW}$ of our sample (see Fig. \ref{fig:mwa_all}) shows that most of the sources have older $t_{MW}$ towards the central region (R$<$0.1R\textsubscript{25}), as expected given the presence (in some cases) of a stellar bulge. In the observed wide age spread ($\Delta(t)\sim 3Gyr$),  $t_{MW}$ can reach up to 7Gyr.
At increasing distances from the centre, the disk component dominates the central bulge. Consistently, at R$>$ 0.2R\textsubscript{25}, a general flattening of $t_{MW}$ is observed, indicating that younger stellar populations have populated these galaxy regions (with typical $t_{MW}\sim4$Gyr).
At larger distances,  R$>$ 0.4R\textsubscript{25}, $t_{MW}$ can follow different patterns: 1) it remains almost flat,  2) it smoothly increases (up to $\sim$ 5Gyr), and 3) it steeply rises to higher age values, as old as the central bulges (up to $\sim$7Gyr), which is indicative of an old stellar populations in the peripheral  regions.


Therefore, we can derive a broad and general result: 
%
The stellar populations in the outermost regions of the disk that are probed by our photometric measurements are older (up to 2 - 3 Gyr) on average than those of the same disk component at  
intermediate galactocentric distances (where they are not contaminated by the bulge component).
We note that the rise is steeper for at least two sources, NGC5194 and NGC4321. The former has an interacting companion (NGC5195), and the latter shows a more extended and dead disk (from 0.8 up to R=R\textsubscript{25}).
We separated the galaxies in our sample into two groups: those with a moderate rising trend in the outer region (NGC3184, NGC0628, NGC5457, NGC4254, NGC4535, and NGC3938), and those with a steeper rising trend (NGC4321 and NGC5194, and even though less extreme, NGC4535).\\

\cite{Casasola17} have measured the scale length (\textit{h}) \footnote{a radial SFR surface density profile fit was performed with a simple exponential curve, $S=S_0exp(-h/h_0)$, where $h_{0}$ is the scale length at the radius $h=0$ \citep{Casasola17}.} in GALEX FUV and NUV (tracing young star emission) and at 3.6$\mu$m emission (mostly related to old stellar populations) for a sample of 18 nearby spiral galaxies. Three of them belong to the sample considered in this paper: NGC0628, NGC5457, and NGC5194. From \cite{Casasola17}, we can quantify the average ratio of the scale lengths over $R_{25}$ for the whole sample within the disk and in the different observed UV and near-IR spectral ranges: \\ \\
$<h_{FUV}/R_{25}> = 0.40$ \\
$<h_{NUV}/R_{25}> = 0.34$ \\
$<h_{3.6 \mu m}/R_{25}> = 0.24$, \\ \\
corresponding to \\ \\
$<h_{FUV}/h_{3.6\mu m}> = 1.67$ \\
$<h_{NUV}/h_{3.6\mu m}> = 1.42$. \\ \\
These scale lengths and ratios of the scale lengths indicate that young stars are on average located in regions of the disk in which the bulge component dominates, as expected. NGC0628 and NGC5457 have ratios of the scale lengths that are consistent with the mean ratios of spiral galaxies: \\ \\ 
$h_{FUV}/h_{3.6\mu m} (NGC0628) = 1.87 (\;\gtrsim 1.67)$  \\ 
$h_{NUV}/h_{3.6\mu m} (NGC0628) = 1.59 (\;\gtrsim 1.42)$ \\ \\
and \\ \\ 
$h_{FUV}/h_{3.6\mu m} (NGC5457) = 1.75 (\; \gtrsim 1.67)$  \\  
$h_{NUV}/h_{3.6\mu m} (NGC5457) = 1.58 (\;\gtrsim 1.42)$. \\ \\
The $h_{FUV}/h_{3.6\mu m}$ and $h_{NUV}/h_{3.6\mu m}$ values are clearly higher than the reference values (1.67 and 1.42) for the galaxies NGC0628 and NGC5457.
This is consistent with our results: We do not find a particular increasing trend for the mass-weighted age within the optical radius for NGC0628 and NGC5457. This scenario is different from that for NGC5194, which shows a different behaviour with respect to the average trend of spiral galaxies within the disk (and the same might hold for NGC4321, which, unfortunately, is missing in the 18 galaxies sample). In this particular case, \\ \\ 
$h_{FUV}/h_{3.6\mu m} (NGC5194) = 1.03 (< 1.67)$  \\ 
$h_{NUV}/h_{3.6\mu m} (NGC5194) = 0.91 (< 1.42)$. \\ \\

The lower than the average values of this ratio for NGC5194 indicate that for this source, the stellar emission from the older populations (whose emission is traced by the 3.6$\mu m$ light) has a flatter and more extended radial profile than the companion spirals. This is in line with our finding for NGC5194 (Fig. \ref{fig:mwa_all}).

We note that older stars 

in the outskirts of the disk 

appear to be counter-intuitive with respect to the predictions of the inside-out quenching mechanism that was widely discussed in the literature to explain the formation of bulges in the local Universe \citep[e.g.][]{Morselli,Tacchella}.
High-$z$ observations found that in the most massive galaxies, star formation is quenched from the inside out, on timescales shorter than one billion years in the inner region and up to a few billion years in the outer disks. These galaxies sustain high star formation activity at large radii while hosting fully grown and already quenched bulges in their cores \citep{Tacchella}.
If this is observed at cosmic noon ($z\sim2$), extrapolations at $z=0$ could naively predict that the progressive quenching of the SFR toward the edges of the disks will lead to an almost passive spheroid as of today \citep{Tacchella}.
\\ In the next sections, we explore NGC4321 in detal to investiage these features. We have a plethora of ancillary data for this galaxy (gas information as well). After this, we explore a possible explanation for the observed quenched rings and the rising age trend, which are visible at large radii in NGC5457 as well.

\subsection{Red and dead stellar ring surrounding NGC4321}
In this section, we further discuss the specific feature observed in NGC4321 (see Fig. \ref{fig:NGC4321}), which is identified as a region of suppressed SFR in the galaxy outskirts, when compared to the expected spatially resolved main sequence.
Before attempting any interpretation, we investigated the significance of the quenched ring to exclude that this is an artefact due to a low S/N in the optical/IR bands. We recall that only the pixels with a 
S/N$>$3
in at least ten photometric bands
were considered for the SED fits. 
However, we note that this condition selected cells with a much higher S/N even in the outskirts of galaxies (e.g. see examples in Appendix \ref{Appendix1}, where S/N$>20$ for all photometric data). 

\begin{figure}[h]
	\centering
		\includegraphics[width = \columnwidth]{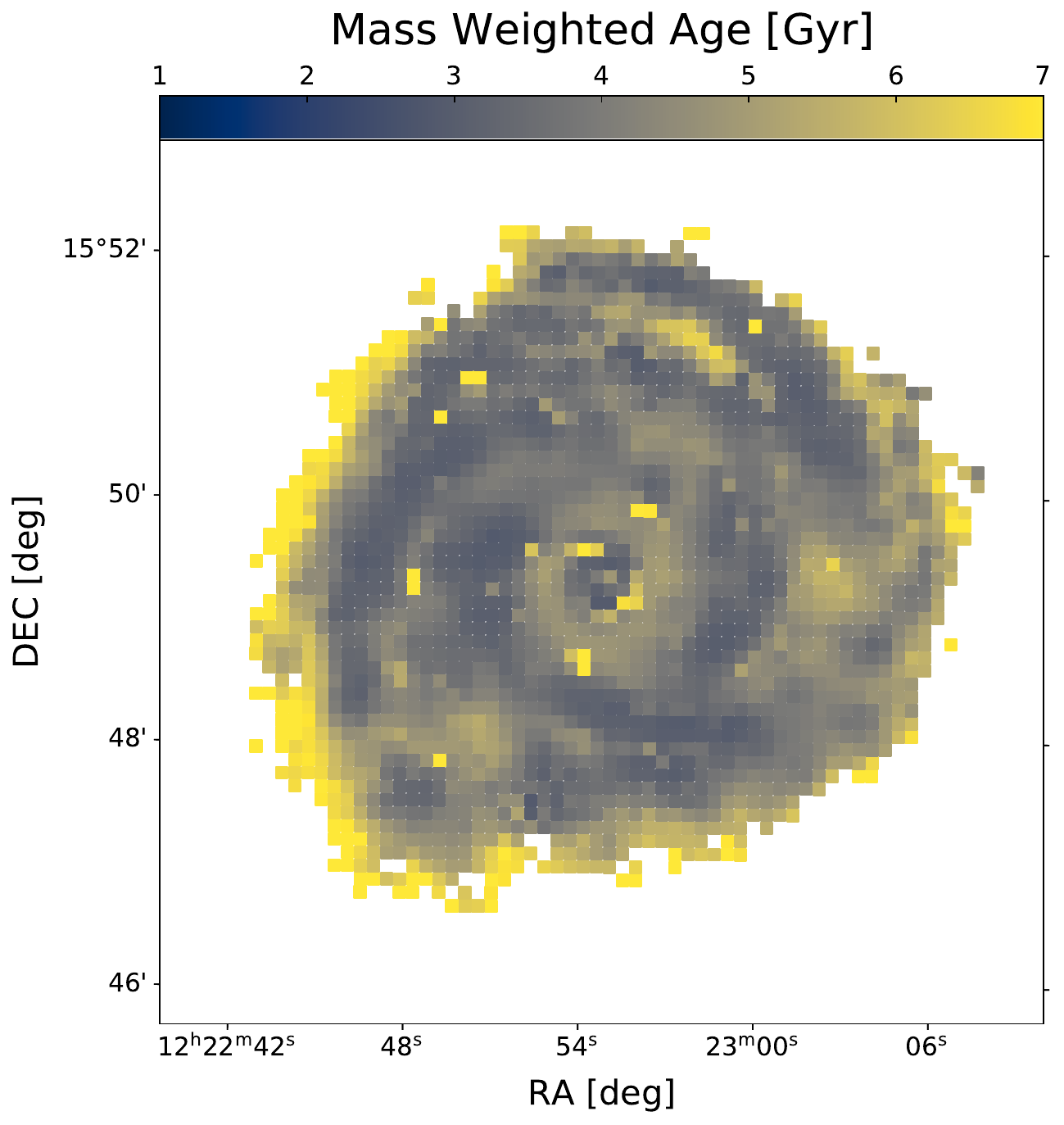}
		\caption{Spatial distribution of the mass weighted age for NGC4321.}
		\label{fig: mwa_4321}
\end{figure}

\begin{figure*}[h]
        \centering
                \includegraphics[width=18cm]{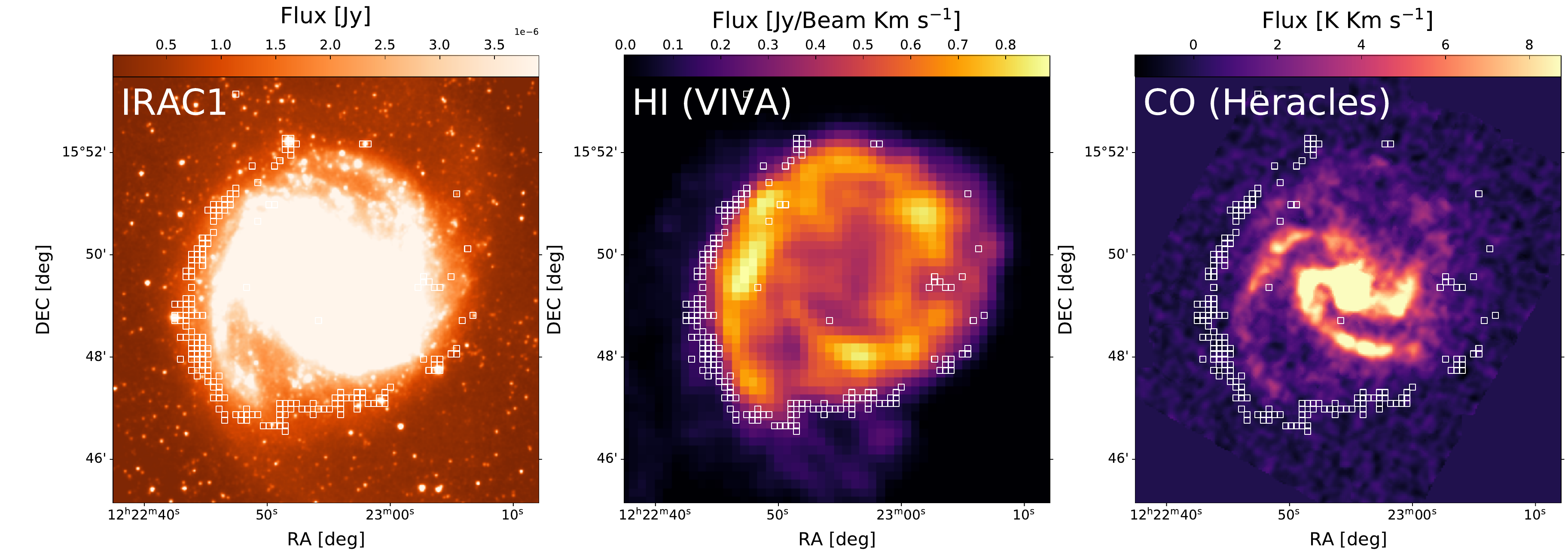}
                \caption{Spatially resolveed properties of the face-on galaxy NGC4321{\it Left panel:} IRAC/Spitzer view of NGC4321. The map at 3.6$\mu$m highlights the diffuse stellar emission around the disk and shows the structure of the spiral arms. {\it Central panel:} HI emission from the public VIVA database. The maps are reported in the same astrometric system. The open squares mark the position of the galaxy cells building the quenched ring, and they have a size of 8"$\times$8". {\it Right panel:} The H2 gas phase  for NGC4321 is traced by the CO emission from the Heracles survey. Each image is presented at the native angular resolution ({\it Spitzer} = 1.6", HI = 15", CO = 11").}
                \label{fig: irac+gas}
\end{figure*}

Fig. \ref{fig: irac+gas} (left panel) shows the infrared emission probed by the IRAC 3.6 $\mu$m  image. This emission is a direct tracer of the stellar mass. NGC4321 shows faint and diffuse starlight emission toward the outskirts, highlighting the existence of old stars in these regions, without evidence of ongoing SF.
Fig. \ref{fig: mwa_4321} shows the mass-weighted ages for NGC4321 cells, 
with an evident  peripheral region around the galaxy with mass-weighted ages of several billion years that correspond to the quenched ring already identified for its suppressed specific SFR.

In Fig. \ref{fig: irac+gas} (left panel) we overplot as small open white squares the positions of the cells that correspond to the passive ring below the MS.  The round distribution of these points at large galactocentric distances further highlights the diffuse stellar emission. Interestingly, the red and dead zone begins where the spiral arms end.
In Fig. \ref{fig: irac+gas} (central and right panels) we further plot the HI and H$_2$ surface densities. The $H_2$  gas phase is traced by the CO emission from the HERACLES survey hrtr \citep{Leroy}, while the HI comes from the public VIVA database \citep{Chung}. The observed CO emission was converted into an H$_2$  gas mass following the prescription described in \cite{Morselli}:
We estimated $\Sigma_{H_2}$ using Eq. 4 of \citet{Leroy}, considering a metallicity-independent conversion factor X$_{CO}$ (X$_{CO}$ = N(H$_2$)/I$_{CO}$, where N(H$_2$) is the H$_2$ column density, and I$_{CO}$ is the line intensity) equal to 2$\times$10$^{20}$ cm$^{-2}$ (K km s$^{-1}$)$^{-1}$  \citep[the typical value for disc galaxies, see e.g.][]{Bolatto}, and a CO line
ratio I$_{CO(2-1)}$/I$_{CO(1-)}$=0.8
\citep[e.g.][]{Leroy,Schruba,Casasola15}.
We divided by a factor 1.36, which was included in Eq. 4 of Leroy et al. (2009), to remove the helium contribution.

The maps are reported in the same astrometric system. Again, the open squares mark the position of the galaxy cells building the quenched ring. For both HI and CO, the S/N is relatively low in the outskirts of the galaxy, and therefore, we stacked the data in all the white squares pertaining to the quenched ring. We then obtained log$(\Sigma_{\rm HI}) = (0.43\pm 0.11) M_{\odot} pc^{-2} $ and log$(\Sigma_{\rm H_2}) = (0.03 \pm 0.65) M_{\odot} pc^{-2} $, which means that there is a significant detection of neutral hydrogen, whereas molecular hydrogen is only marginally detected, but with a large uncertainty. A low  H$_2$ surface density is also to be expected because HI is well below the $\sim 10\,M_\odot$ pc$^{-2}$ threshold for the production of molecular hydrogen \citep{Bigiel}.
Thus, in the quenched ring, H$_2$ appears to be lacking, while the atomic HI is still present, as is often observed in local spiral galaxies \citep{Casasola17}. 

\begin{figure*}[htpb]
        \centering
                \includegraphics[width =19cm]{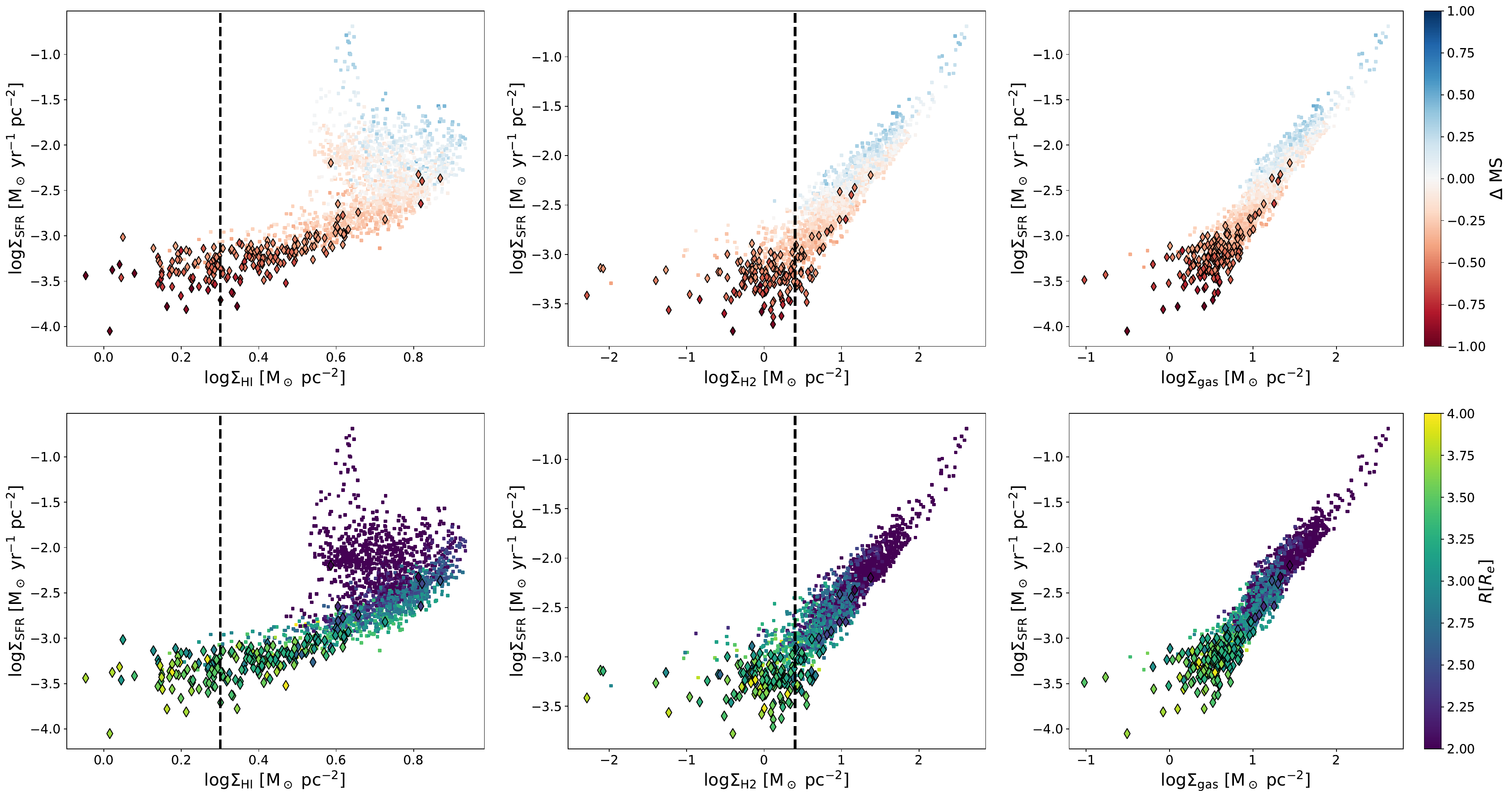}
                \caption{Spatially resolved Kennicutt-Schimdt relation for NGC4321. The surface density of SFR ($\Sigma_{SFR}$) is reported as a function of  $\Sigma_{HI}$ (left panel), $\Sigma_{H2}$ (central panel), and the total (HI+H2) gas surface density (right panel). 
The three upper panels are colour-coded as a function of the distance to the MS. The lower panels indicate the galacto-centric distance of each physical galaxy cell.
Diamonds correspond to regions associated with the quenched ring.}
                \label{fig: KS}
\end{figure*}

Quantitatively, we report in Fig. \ref{fig: KS} the spatially resolved Kennicutt-Schmidt relation for NGC4321, that is, we compare the surface density of SFR ($\Sigma_{SFR}$) versus   $\Sigma_{HI}$ (left panel), $\Sigma_{H_2}$ (central panel) and the total (HI+H$_2$) gas surface density (right panel). 
As already mentioned, the H$_2$  signal is almost undetected in these very external regions, with most of the marginal signal falling below the sensitivity limit of the HERACLES survey (marked by the vertical dashed line in the central panels).

A low $\Sigma_{HI}$ like this, associated with a paucity of $H_2$, in addition to a distribution of an old stellar population, indicates that "starvation" has started at the outer edge of the galaxy as a result of insufficient gas supply, whereas star formation is sill going on in the inner regions.

\begin{figure}[htpb]
        \centering
                \includegraphics[width = \columnwidth]{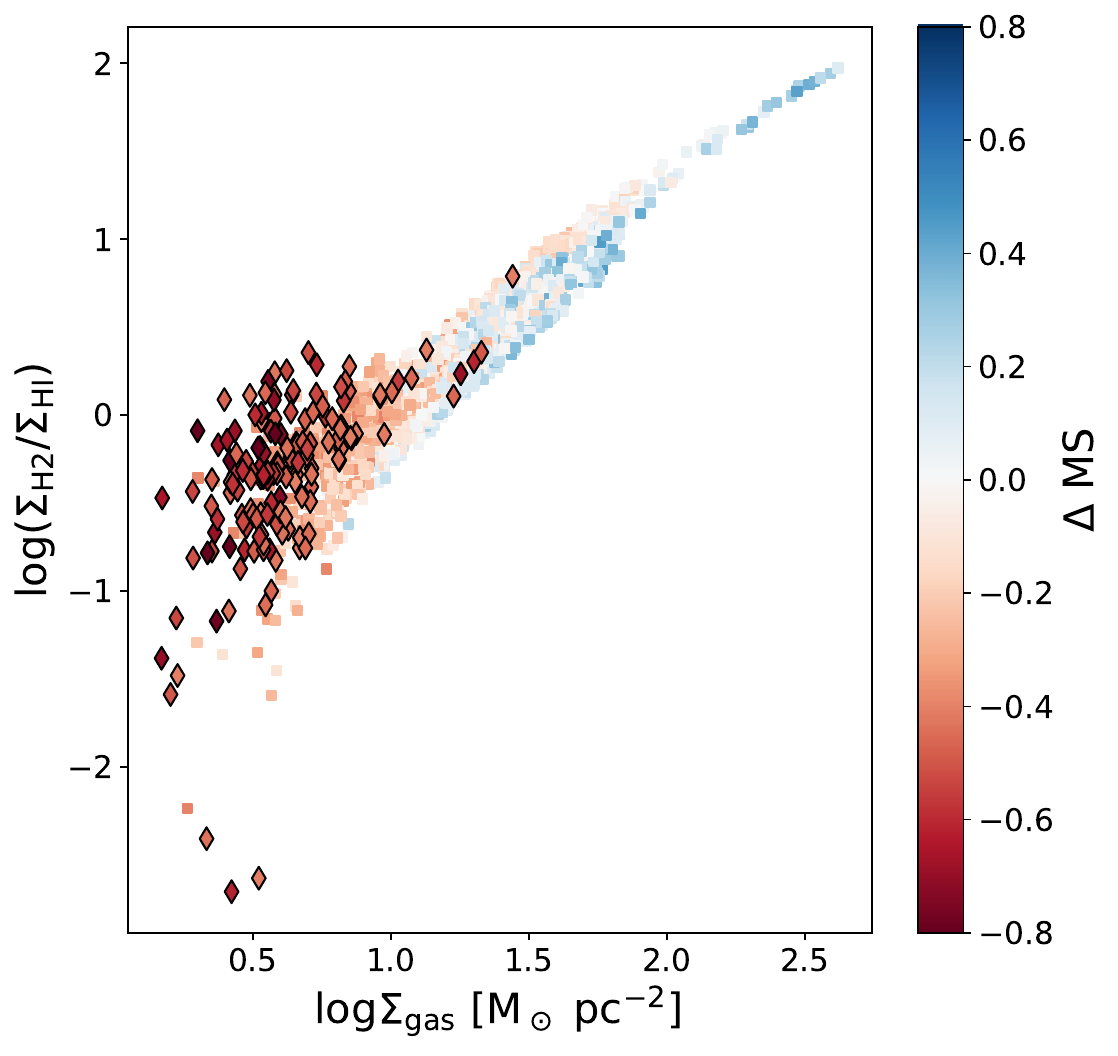}
                \caption{Spatially resolved ratio of the surface densities of H2/HI, $\Sigma_{H2/HI}$, shown as a function of the total gas density.  The colour code marks  the distance to the MS. Diamonds correspond to regions associated with the quenched ring.}
                \label{fig: H2/HI}
\end{figure}

This is also supported by  Fig. \ref{fig: H2/HI}, where the ratio of the surface densities of H$_2$ /HI ($\Sigma_{H_2 /HI}$) is shown as a function of the total gas density. The quenched ring regions (represented as diamonds) follow the general trend observed for local spiral galaxies \citep[see Fig. 8 in][]{Morselli}, being well below the MS.


The direct evidence is that the quenched ring contains insufficient gas to form enough H$_2$ to sustain star formation.  This may have two reasons: i) either gas has been removed from the ring, or ii) after consumption to form stars, it has not been replaced by fresh gas from the circumgalactic medium (CGM).

NGC4321 is located in the Virgo cluster, and it therefore moves through its intracluster medium (ICM), and ram pressure stripping from the diffuse plasma is well known to deplete the gas from galaxies within a cluster, quenching the ongoing SFR \citep[e.g.][]{Poggianti,Poggianti1,Boselli}. 
For example, in the case of the cluster galaxy IC3476, Boselli et al. (2021) showed that the IGM-ISM interaction causes the overall rise in the star formation activity within the galaxy (as testified by its multiple giant HII regions) and the sudden drop in star formation at the edge of the disk 
(some $\sim 50$ Myr ago), which is completely quenched, having been stripped of its gas content.
Therefore, it is natural to entertain the possibility that while moving through the Virgo cluster, NGC4321 received the same treatment.
However, other sources from the \citet{Enia} sample (in particular, NGC5194; see Fig. \ref{fig:mwa_all} show trends very similar to that of NGC4321, including  the connection between  quenched stellar populations and low gas surface density. However, these objects are not associated with any cluster.

The second alternative may apply to them, namely that the CGM has ceased to supply fresh gas to the galaxy. According to cosmological models, to sustain their SFR across cosmic  time, galaxies are fed by cold gas streams \citep{Dekel}. The gas appears to be accreted primarily in a co-planar, co-rotating  fashion \citep[e.g.][]{Bouche}, hence with increasing angular momentum. On this basis, \cite{PengRenzini} have argued that a time may come in the life of a galaxy when the residual CGM is left with too high specific angular momentum to be accreted effectively, thus leading to starvation and quenching, starting from the outer edge of the disk.

\section{Summary and conclusions}
We used the BAGPIPES code to perform a spatially resolved SED-fitting in a sample of eight local grand-design spiral galaxies, drawn from  the DustPedia database. Accounting for photometric information from the UV up to the far-infrared, we produced maps of stellar mass density, star formation rate density, and distance from the MS, using as reference the relation derived by \citet{Enia}.
Focusing on the analysis of the SFHs in each galaxy region, which were conservatively assumed in a parametric form (delayed-exponential models), we studied the distribution of the mass-weighted ages ($t_{MW}$ as indicator of the main epoch of the stellar build up) at different galacto-centric distances, and compared the average trends of the whole galaxy sample.



The average distribution of the mass-weighted age of our sample shows two different behaviours from the centre toward the peripheral regions: 
$t_{MW}$ indicates older populations in the central region ($t_{MW}$ up to $\sim$7Gyr at $R_{25}<0.3$), consistent with the presence of a bulge for various systems; and at larger distances, the disks are dominated by younger regions that are still forming stars ($t_{MW}\sim$4Gyr).

If an average mild rise of $t_{MW}$ is seen at $R_{25}>0.3$, mostly consistent with a flat relation, NGC4321 and NGC5194 instead present a dramatic steep increase of $t_{MW}$ that reaches levels as high as to those of the bulges.
The galaxy loci that are characterised by old stellar populations like this belong to the  external regions of the disk and lie immediately beyond the influence of the spiral arms. Their specific star formation rates fall well below the corresponding spatially resolved MS. We thus call them {\it quenched rings}.

We discussed the origin of these peculiar feature by exploiting further physical information about the  different gas phases (HI and H\textsubscript{2}) available in particular for NGC4321, the galaxy with the most evident quenched ring. One possibility is that the outer ring has been depleted in gas by ram pressure stripping, and indeed, NGC4321 is a member of the Virgo cluster. However, NGC5194 is not associated with any cluster, and ram pressure stripping could not be invoked for its quenched ring. Alternatively, accretion of cold gas may have ceased or diminished in this case because the angular momentum of the remaining circumgalactic medium is too high to be incorporated into the galaxy.

We plan to expand this type of study to a statistically more significant sample from the DustPedia database. 


\begin{acknowledgements}
We thank the referee for valuable suggestions that improved the paper.
G.R., A.E. and L.B. acknowledge the support from grant PRIN MIUR 2017 - 20173ML3WW_001. 
GR also acknowledges INAF under the Large Grant 2022 funding scheme (project "MeerKAT and LOFAR Team up: a Unique Radio Window on Galaxy/AGN co-Evolution).
VC acknowledges funding from the INAF Mini Grant 2022 program “Face-to-Face with the Local Universe: ISM’s Empowerment (LOCAL)”.

\end{acknowledgements}

%

%


\bibliographystyle{aa}
\bibliography{biblio}

\appendix
\section{BAGPIPES SED plots}
\label{Appendix1}
In this section, three SED plots are reported. Each plot corresponds to an aperture located at a different distance from the centre of the galaxy NGC4321.

\begin{figure}[htpb]
        \centering
                \includegraphics[width = \columnwidth]{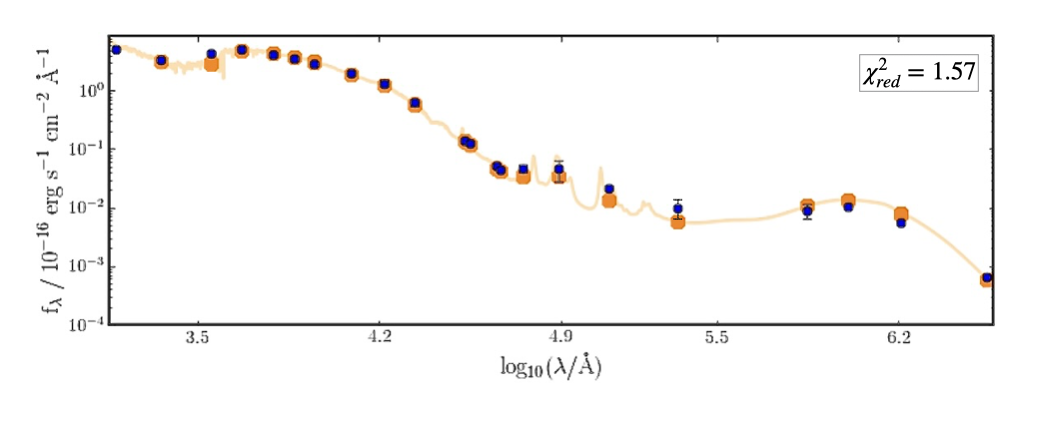}
                \caption{BAGPIPES SED-fitting plot for an aperture located in the centre of NGC4321 (aperture A in \ref{fig:NGC4321}). The data are shown in blue, and the posterior distribution, given by the difference of the $16^{th}$ and the $84^{th}$ percentiles, is shown in orange.
  The error bars are reported for the observed data. When they not visible, they are smaller than the corresponding symbols.}
        \label{center}

\end{figure}

\begin{figure}[htpb]
        \centering
                \includegraphics[width = \columnwidth]{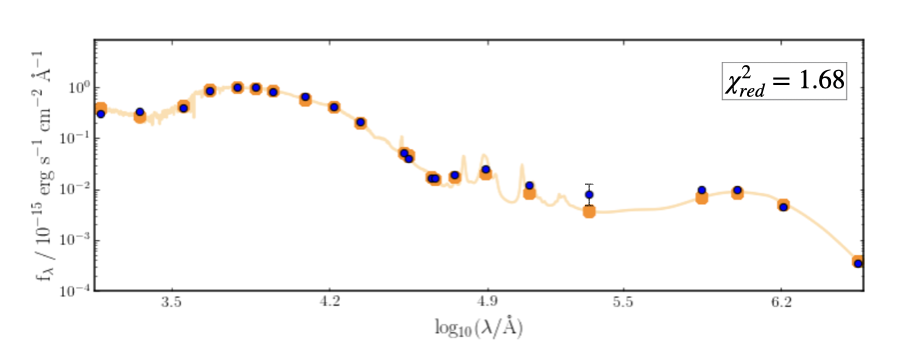}
                \caption{Same as \ref{center} for an aperture located at 0.5$R_{25}$ of NGC4321 (aperture B in \ref{fig:NGC4321}).}
                
\end{figure}

\begin{figure}[htpb]
        \centering
                \includegraphics[width = \columnwidth]{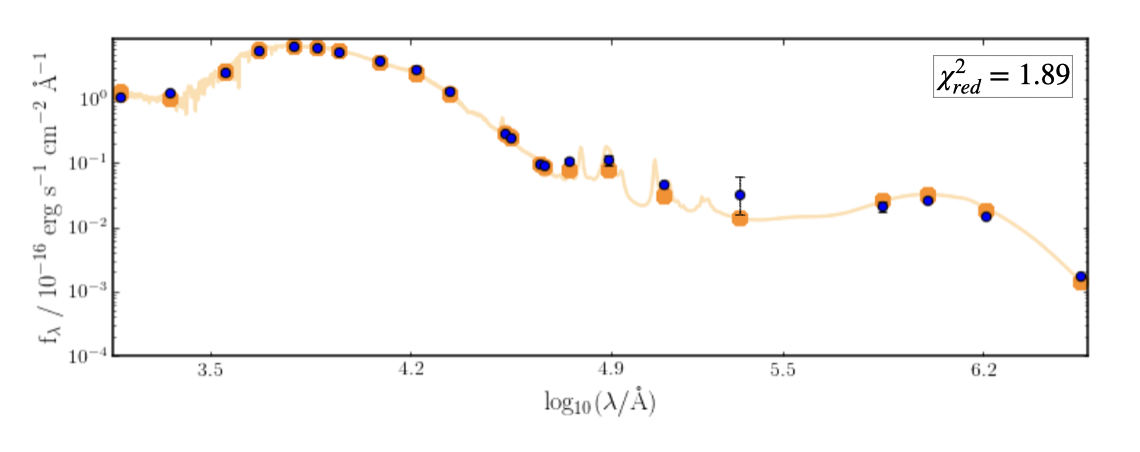}
                \caption{Same as \ref{center} for an aperture located at $R_{25}$ of NGC4321 (aperture C in \ref{fig:NGC4321}).}
                
\end{figure}

 
\section{Comparison with Enia et al. (2020)}
\label{Comparison}

\subsection{Setting the frame: Photometry and MAGPHYS results}
Here, we briefly set the frame for the BAGPIPES SED-fitting procedure by reporting the approach adopted and widely discussed in \citet{Enia}. The strategy consists of three steps: 
\\- The background-subtracted maps were degraded to the 
SPIRE350 PSF ( 8\textsuperscript{$\prime\prime$} ,the worst one in the available photometric bands).
\\ - In a grid of 8\textsuperscript{$\prime\prime$} \textit{x} 8\textsuperscript{$\prime\prime$} cells, we measure the flux at each wavelength in each aperture; (therefore, there are variations in the physical scale length among galaxies, as shown in \ref{tab:1});
\\ - By applying the SED-fitting technique to the given photometry, we can estimate the physical properties of each cell. To perform SED fitting, \citet{Enia} used the publicly available
code MAGPHYS (\citet{daCunha}). Here, we briefly report the main results.
 \\ 
 The values of M\textsubscript{$\star$} were obtained by fitting the 
 DustPedia photometry with MAGPHYS, while SFRs were obtained as $SFR = SFR_{UV} + SFR_{IR}$ where $SFR_{UV}/(M\textsubscript{$\odot$}/yr) = 0.88 \times 10\textsuperscript{-28}\, \textit{L}_\nu/\textit{L}_{\odot}$
with \textit{L}\textsubscript{$\nu$}, in \textit{erg/s/Hz}, as the luminosity per unit frequency evaluated at 150 $n$m \citep[from][]{Bell} and $SFR_{IR} = 2.64 \times 10\textsuperscript{-44}\, \textit{L}_{IR}$ with \textit{L}\textsubscript{IR}, in \textit{erg/s}, as the luminosity evaluated from the SED fit between 8$\mu$m and 1000$\mu$m \citep[][]{Kennicutt}. For each source, it was possible to plot the maps of stellar mass, star formation rate, the distribution of the cells in the $\log\Sigma\textsubscript{$\star$}$-$\log\Sigma\textsubscript{SFR}$ plane (i.e. the spatially resolved MS)and $\Delta\textsubscript{MS}$ evaluated as the perpendicular distance from the MS relation of a point in the $\log\Sigma\textsubscript{$\star$}$-$\log\Sigma\textsubscript{SFR}$ plane. By adopting the EMCEE \citep[][]{Foreman-Mackey} fitting tool, \citet{Enia} performed the fit to compute the MS relation. The fit, a log-linear relation, described as $\log\Sigma\textsubscript{SFR} = m\,\log\Sigma\textsubscript{$\star$} + q$
gave \textit{m} = 0.82 $\pm$ 0.12 and \textit{q} = -8.69 $\pm$ 0.97. The M\textsubscript{$\star$} and SFR MAGPHYS outputs were compared with the values obtained from this work. 
  
\subsection{MAGPHYS and BAGPIPES initial mass function}
Before we describe the  details, we briefly discuss in this section potential offsets in terms of the initial fass Function (IMF) adopted in the MAGPHYS code \citep[a log-normal IMF,][]{Chabrier} and in the BAGPIPES code (a broken power law IMF, \citet{Kroupa}), which may cause residual inconsistency in stellar mass and star-formation rate results with respect to the previous work \citep[][]{Enia}. For a straightforward comparison of the two codes, we then converted all outputs into the Chabrier IMF \citet{Chabrier}. As an example, in \ref{fig:Kroupa&Chabrier}, we show the direct correlation between the stellar mass and SFR from \citet{Enia} and those obtained in this work: the stellar masses are almost consistent, with an average offset of $\sim0.05$ dex  and a scatter of 0.29 dex. In the SFR outputs, the discrepancy between the results from the two SED-fitting codes is only slightly larger: The SFRs from MAGPHYS are $\sim$0.07 dex higher on average than those from BAGPIPES, with a scatter of $\sim$0.30 dex. We can then safely assume that the two codes provide consistent results, in particular, for the stellar mass. The agreement for the SFR from the two SED-fitting procedures is also within the typical uncertainties from different SFR tracers.
We further note that the SFR, averaged over the past 100Myr, derived from an SED-fitting adopting a parametric SFH, can be biased (overestimated or underestimated) because it is sensitive to instantaneous bursts \citep[see][]{Haskell}. For these reasons, we used the empirical SFR obtained by combining the UV and far-IR luminosities (see Section \ref{SFH}).

\begin{figure}[h]
        \centering
                \includegraphics[width=\linewidth,keepaspectratio]{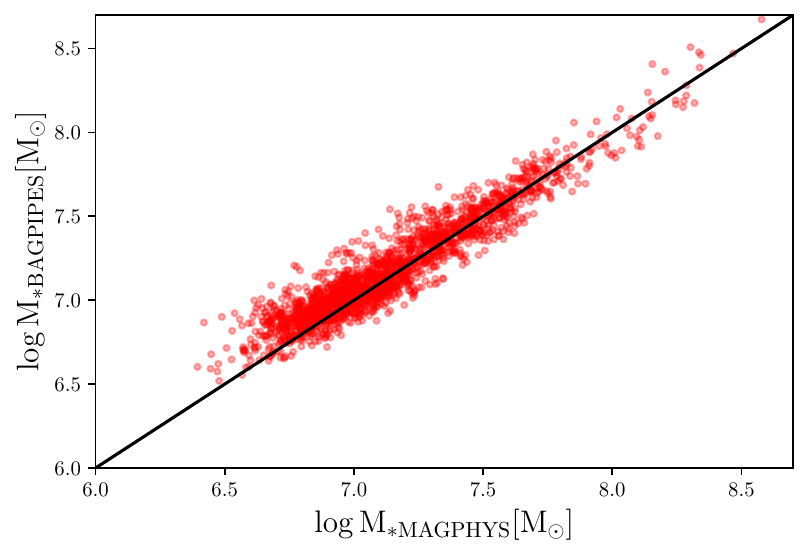}
                \caption{Stellar mass comparison from MAGPHYS and BAGPIPES.}
\end{figure}

\begin{figure}[h]
                \centering
                \includegraphics[width=\linewidth,keepaspectratio]{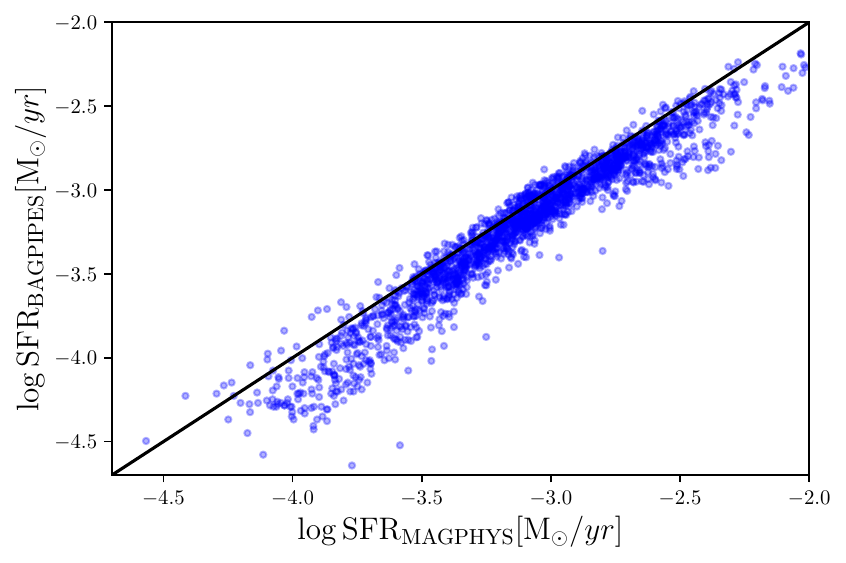}
                \caption{Star formation rate comparison from MAGPHYS and BAGPIPES.}
                \label{fig:Kroupa&Chabrier}
\end{figure}

\newpage
\section{Galaxy-by-galaxy outputs}
\label{galaxy-outputs}
In this section, we present the outputs for each galaxy as shown for NGC4321 in Sec. \ref{SFH}.  
 \begin{figure*}[htpb]
        \centering
        \begin{minipage}{1\columnwidth}
                \centering
                \includegraphics[width=\textwidth]{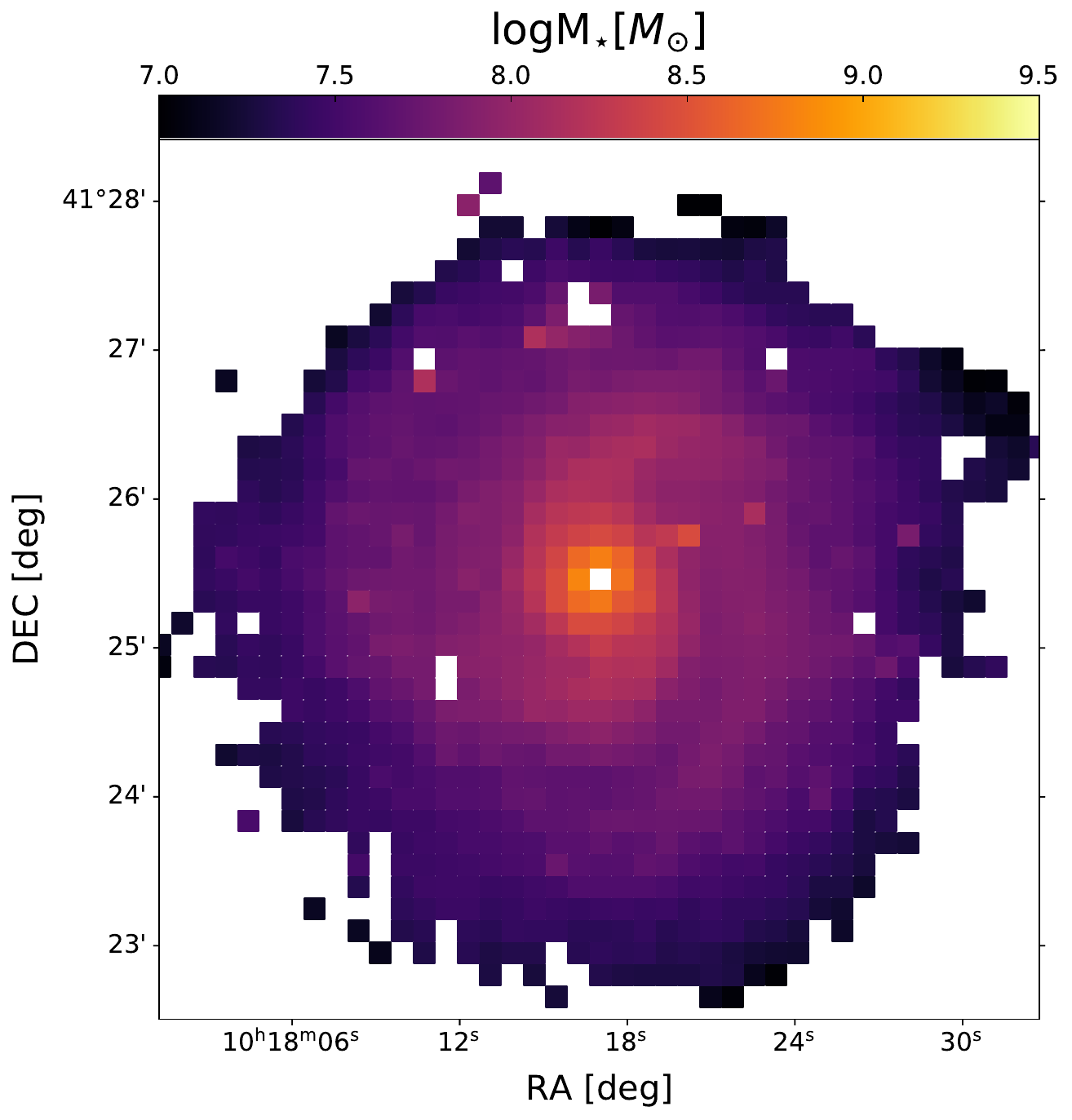}
                
                \label{label1}
        \end{minipage}%
        \begin{minipage}{1\columnwidth}
                \centering
                \includegraphics[width=\textwidth]{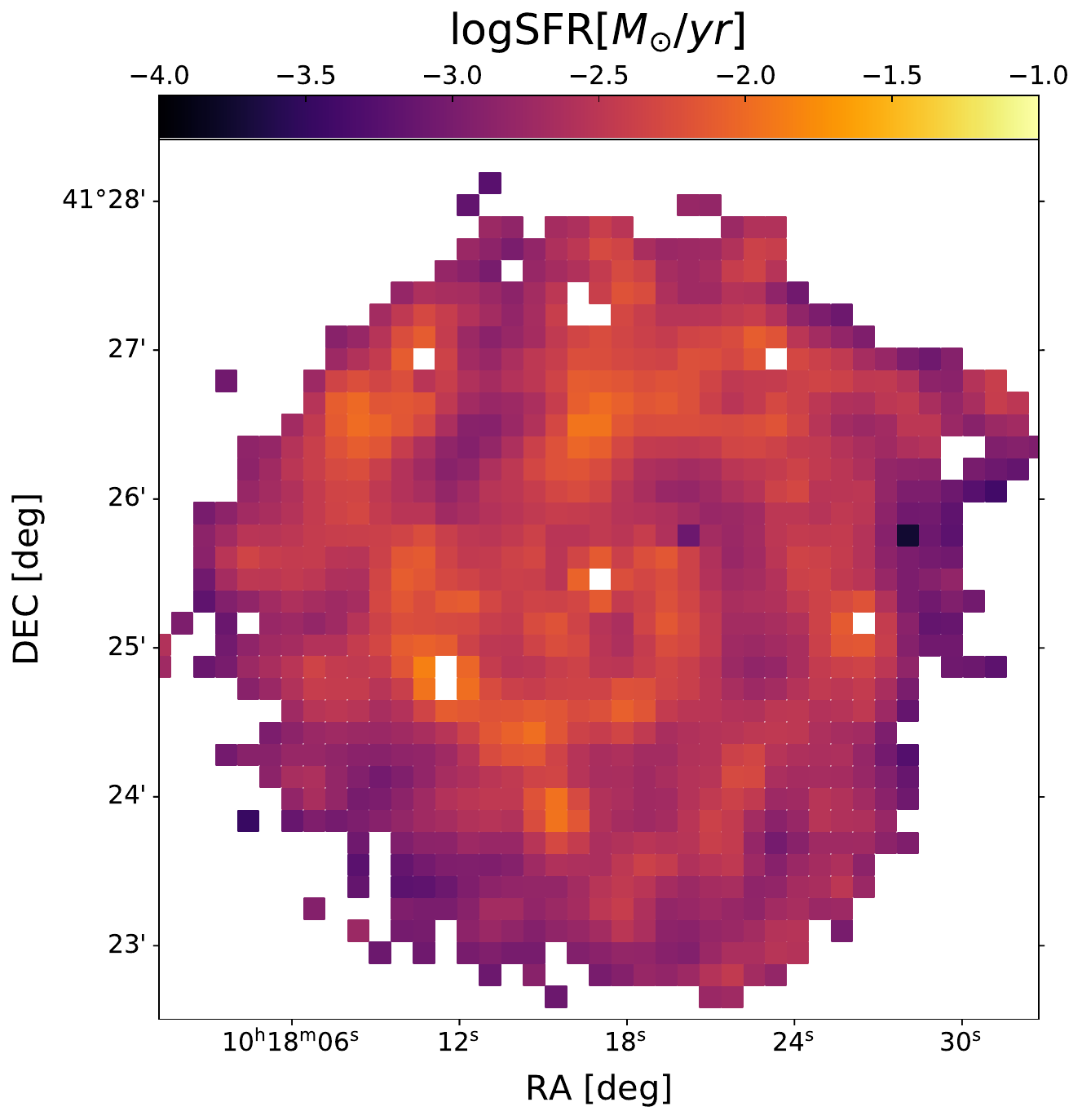}
                
                \label{label2}
        \end{minipage}
 \end{figure*}

 \begin{figure*}[htpb]
        \centering
        \begin{minipage}{1\columnwidth}
                \centering
                \includegraphics[width=\textwidth]{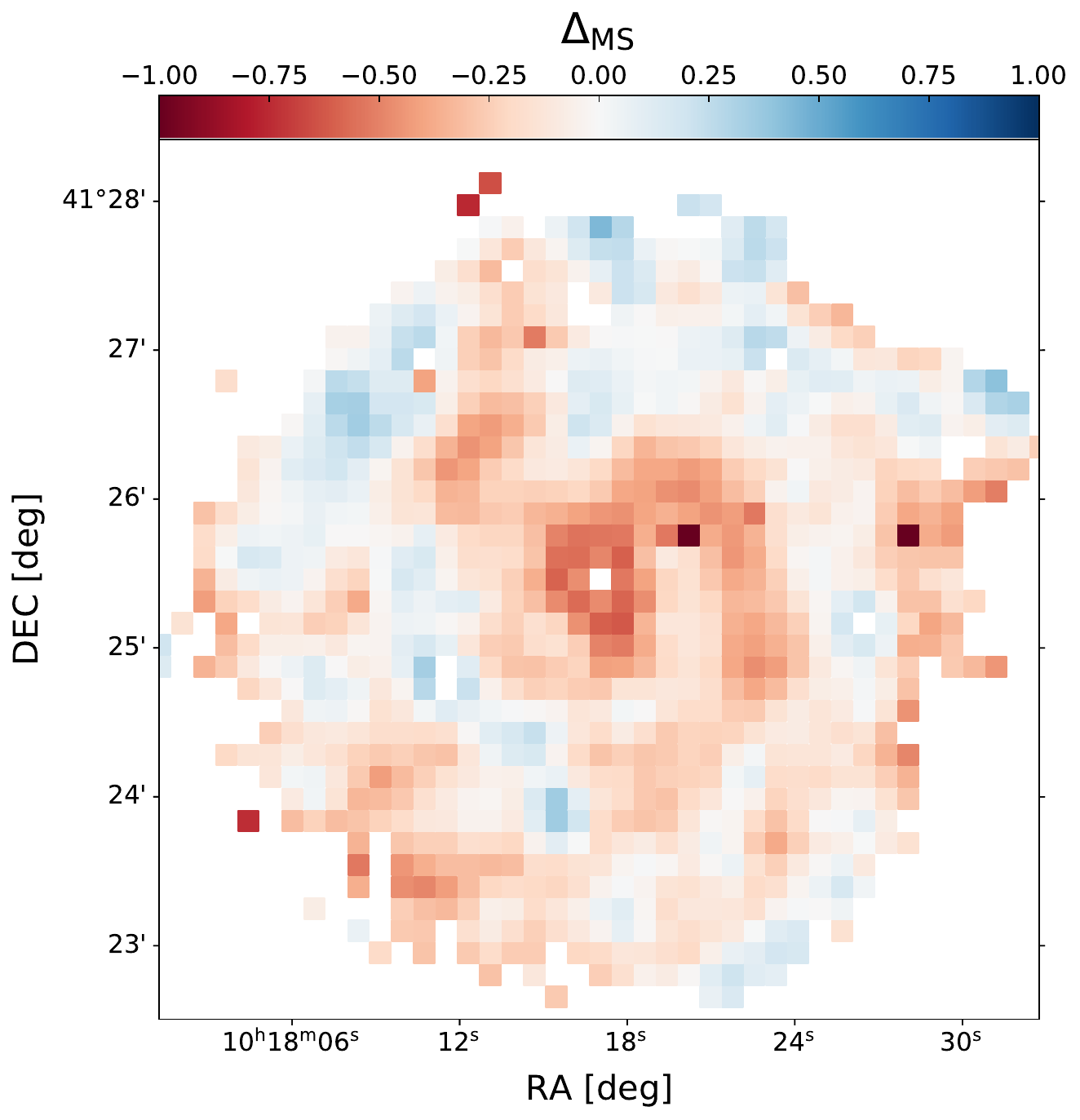}
                
                \label{label3}
        \end{minipage}%
        \begin{minipage}{1\columnwidth}
                \centering
                \includegraphics[width=\textwidth]{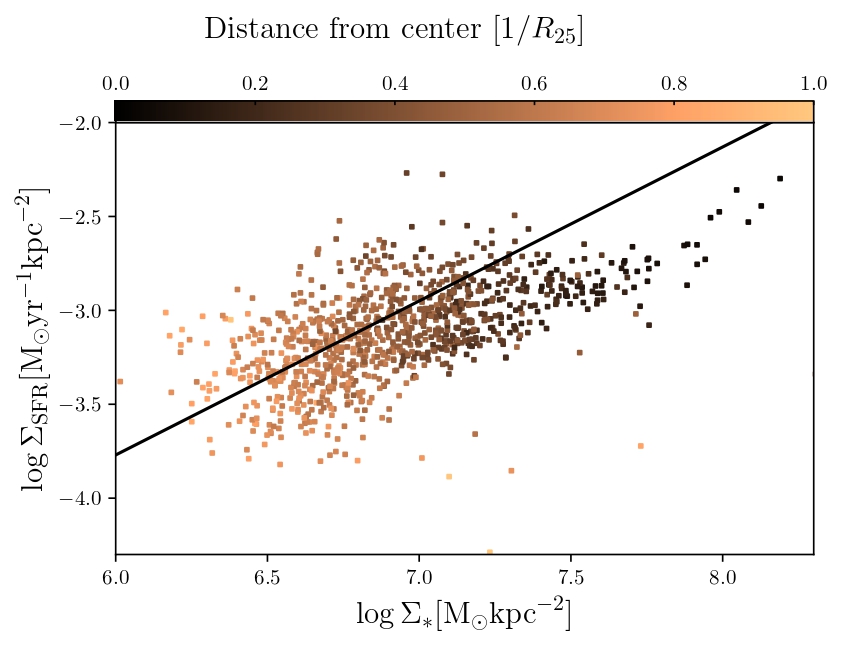}
        
                \label{label4}
        \end{minipage}
        \caption{Same as Fig.\ref{fig:NGC4321}, for NGC3184}
 \end{figure*}


\begin{figure*}[htpb]
        \centering
        \begin{minipage}{1\columnwidth}
                \centering
                \includegraphics[width=\textwidth]{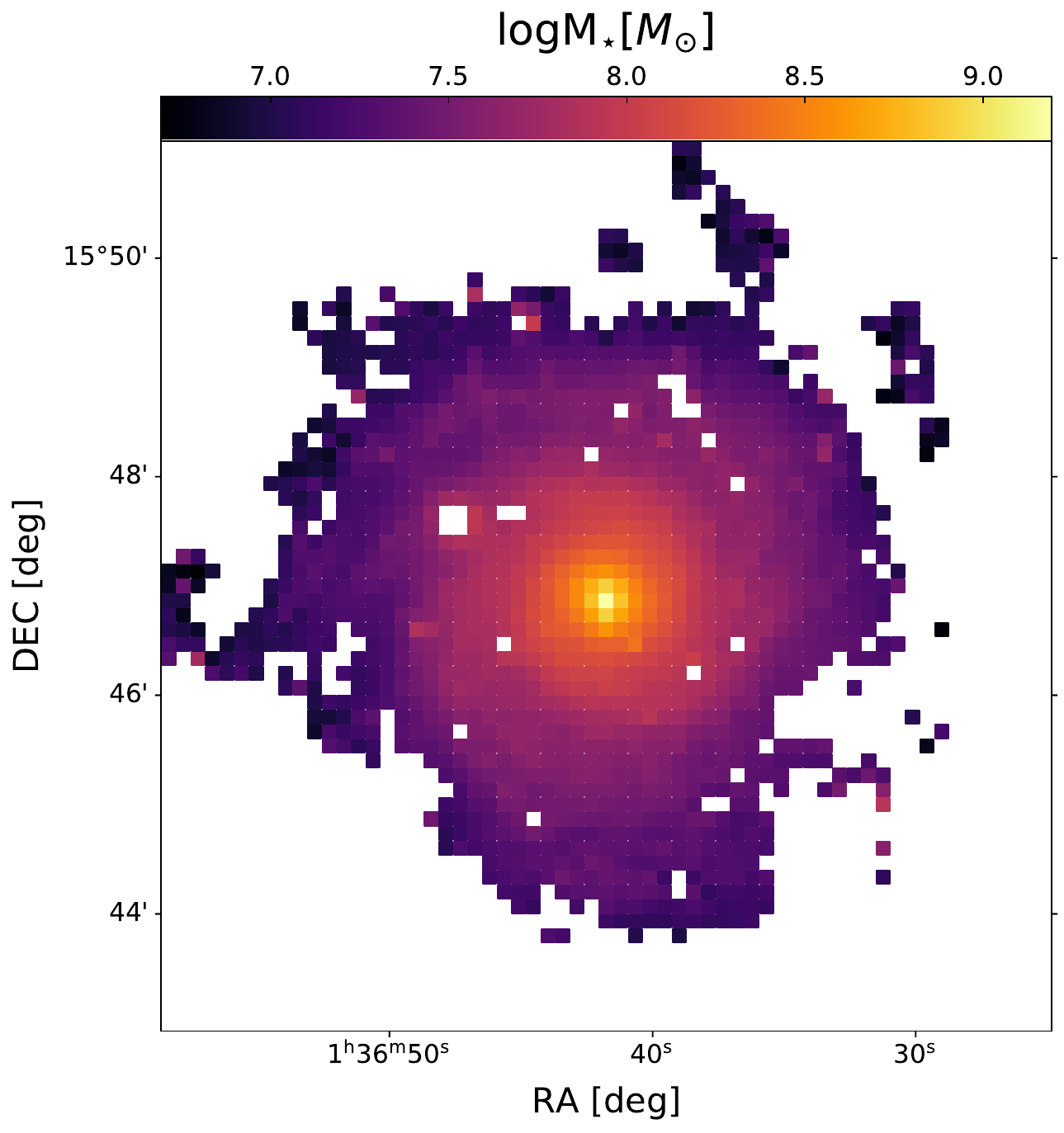}
                
                \label{label1}
        \end{minipage}%
        \begin{minipage}{1\columnwidth}
                \centering
                \includegraphics[width=\textwidth]{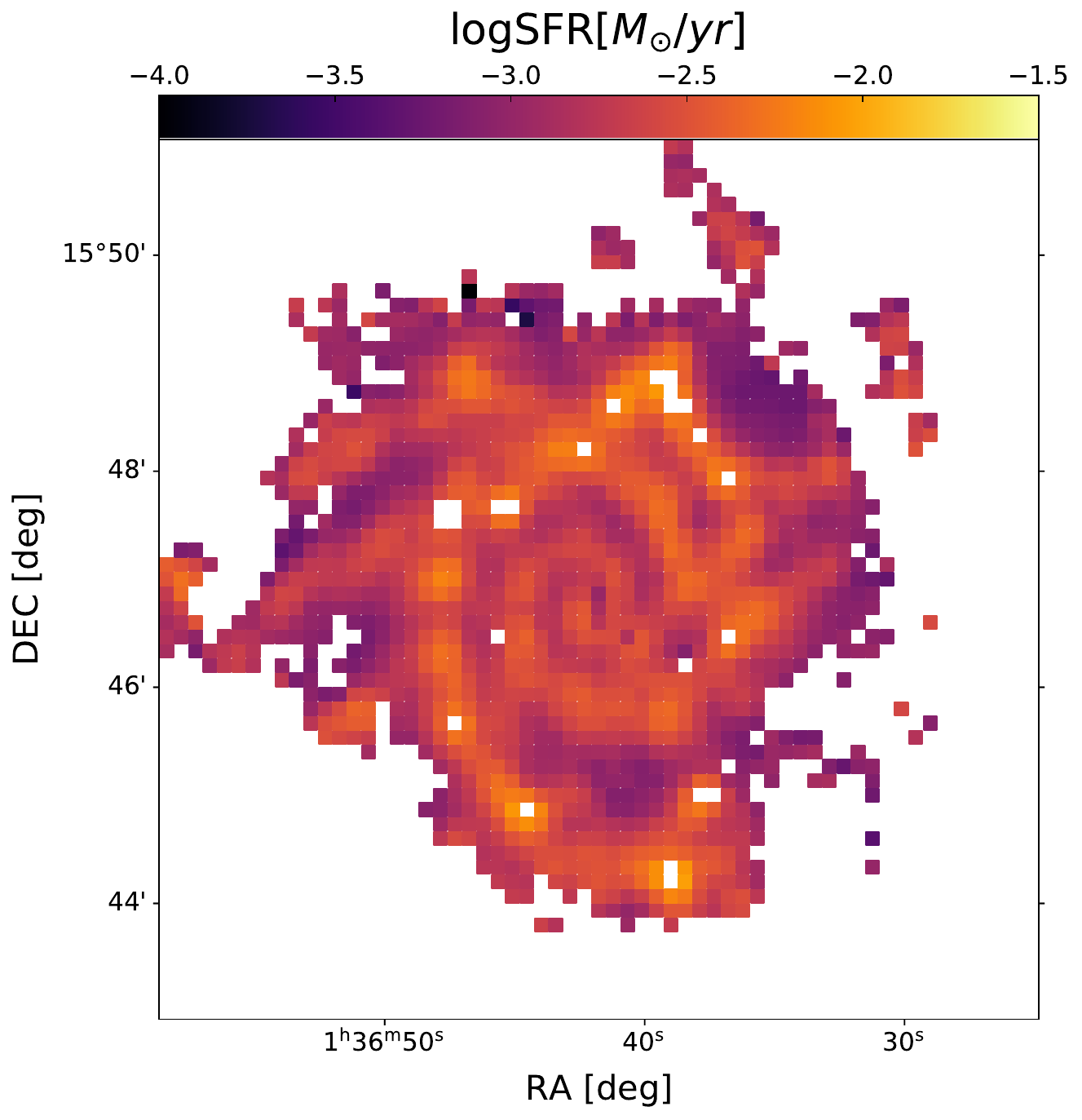}
                
                \label{label2}
        \end{minipage}
\end{figure*}

\begin{figure*}[htpb]
        \centering
        \begin{minipage}{1\columnwidth}
                \centering
                \includegraphics[width=\textwidth]{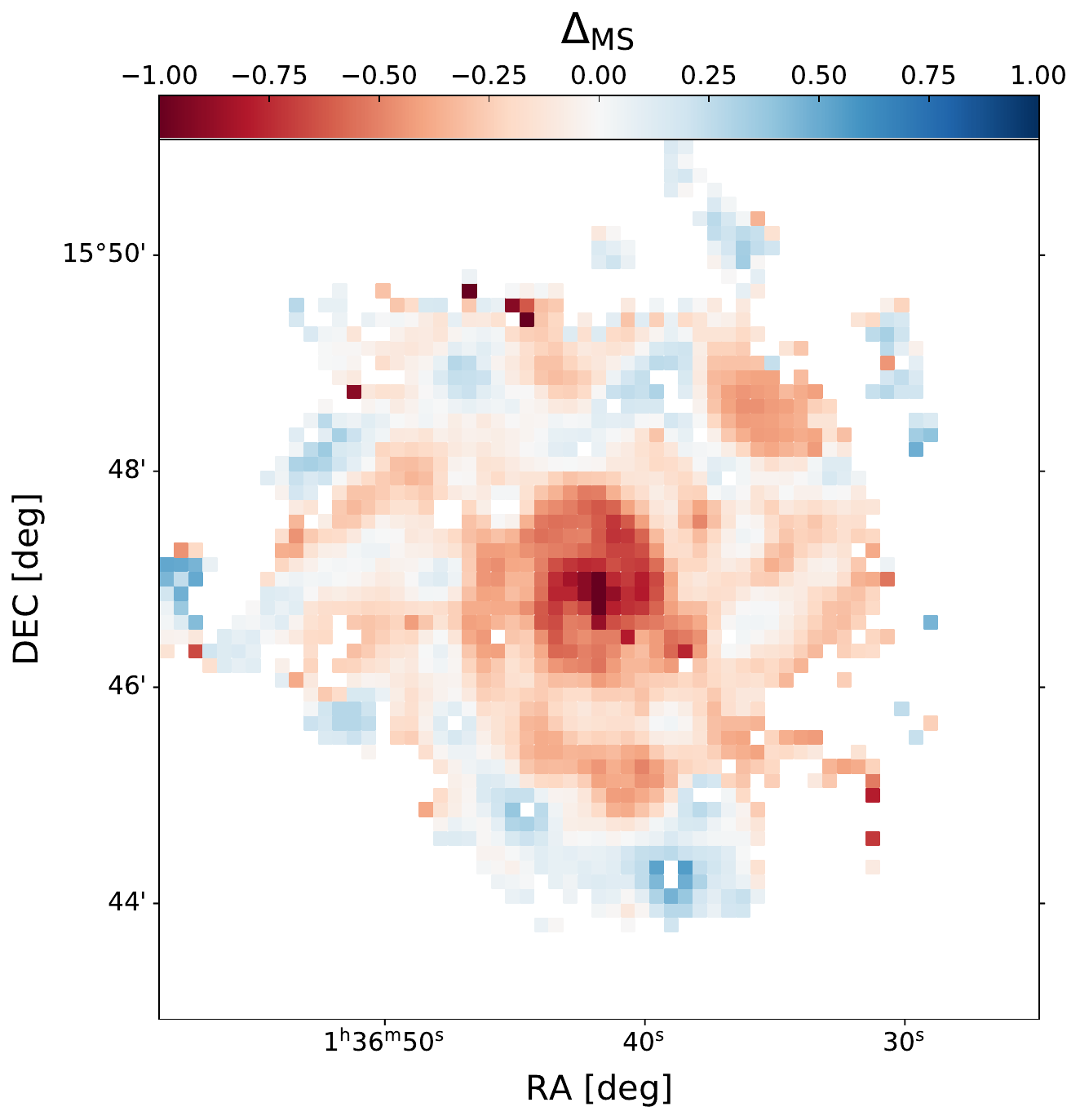}
                
                \label{label3}
        \end{minipage}%
        \begin{minipage}{1\columnwidth}
                \centering
                \includegraphics[width=\textwidth]{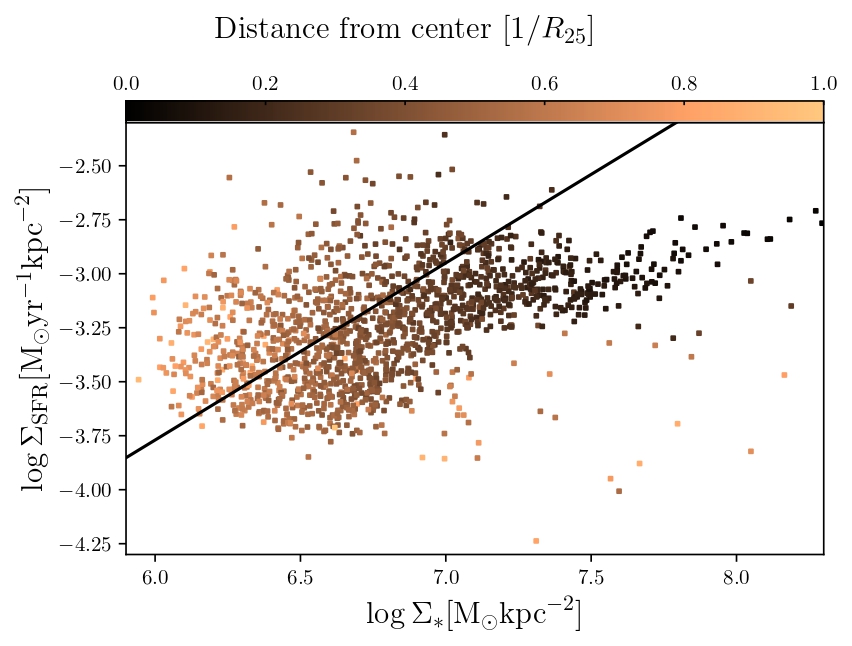}
        
                \label{label4}
        \end{minipage}
        \caption{Same as Fig.\ref{fig:NGC4321} for NGC0628}
\end{figure*}

\begin{figure*}[htpb]
        \centering
        \begin{minipage}{1\columnwidth}
                \centering
                \includegraphics[width=\textwidth]{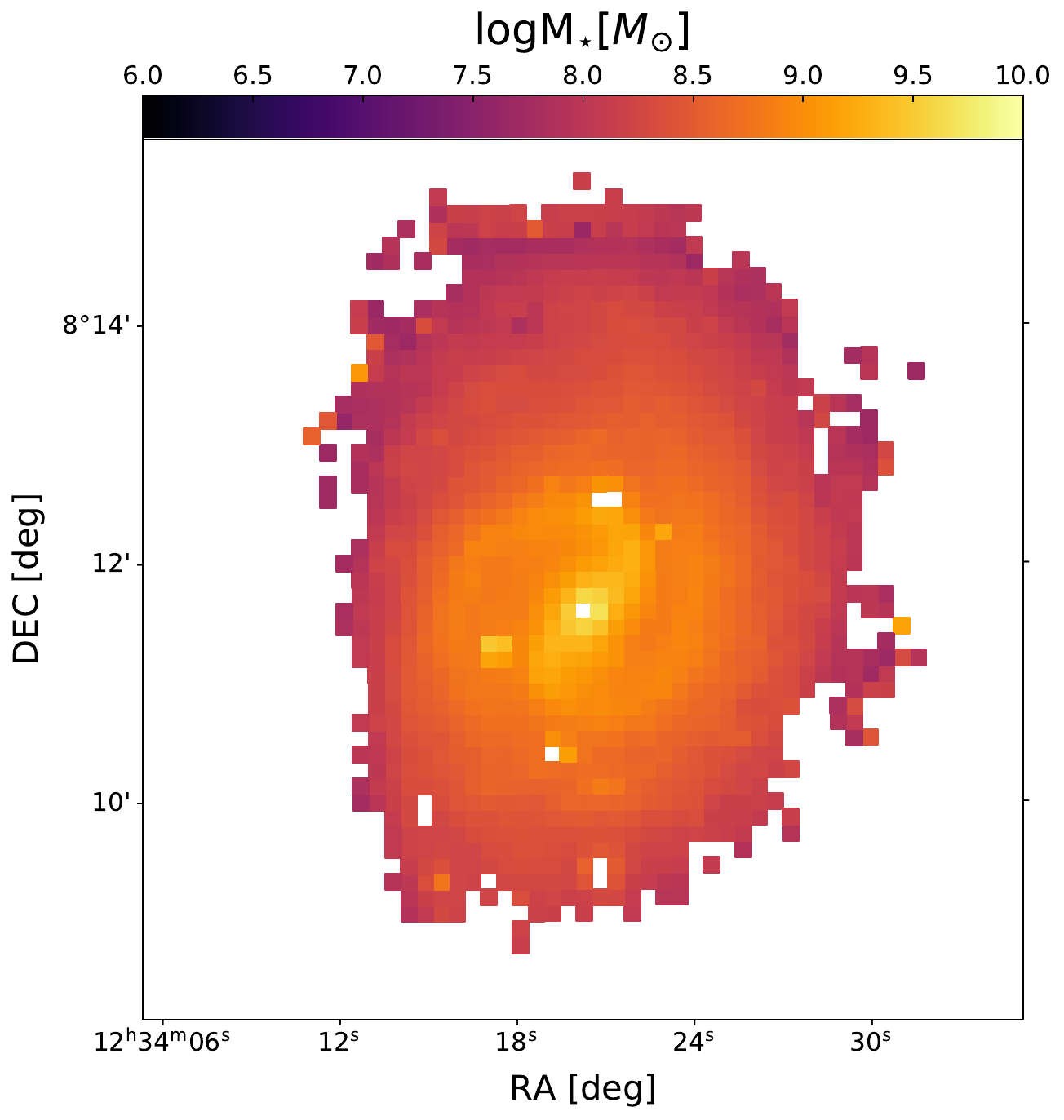}
                
                \label{label1}
        \end{minipage}%
        \begin{minipage}{1\columnwidth}
                \centering
                \includegraphics[width=\textwidth]{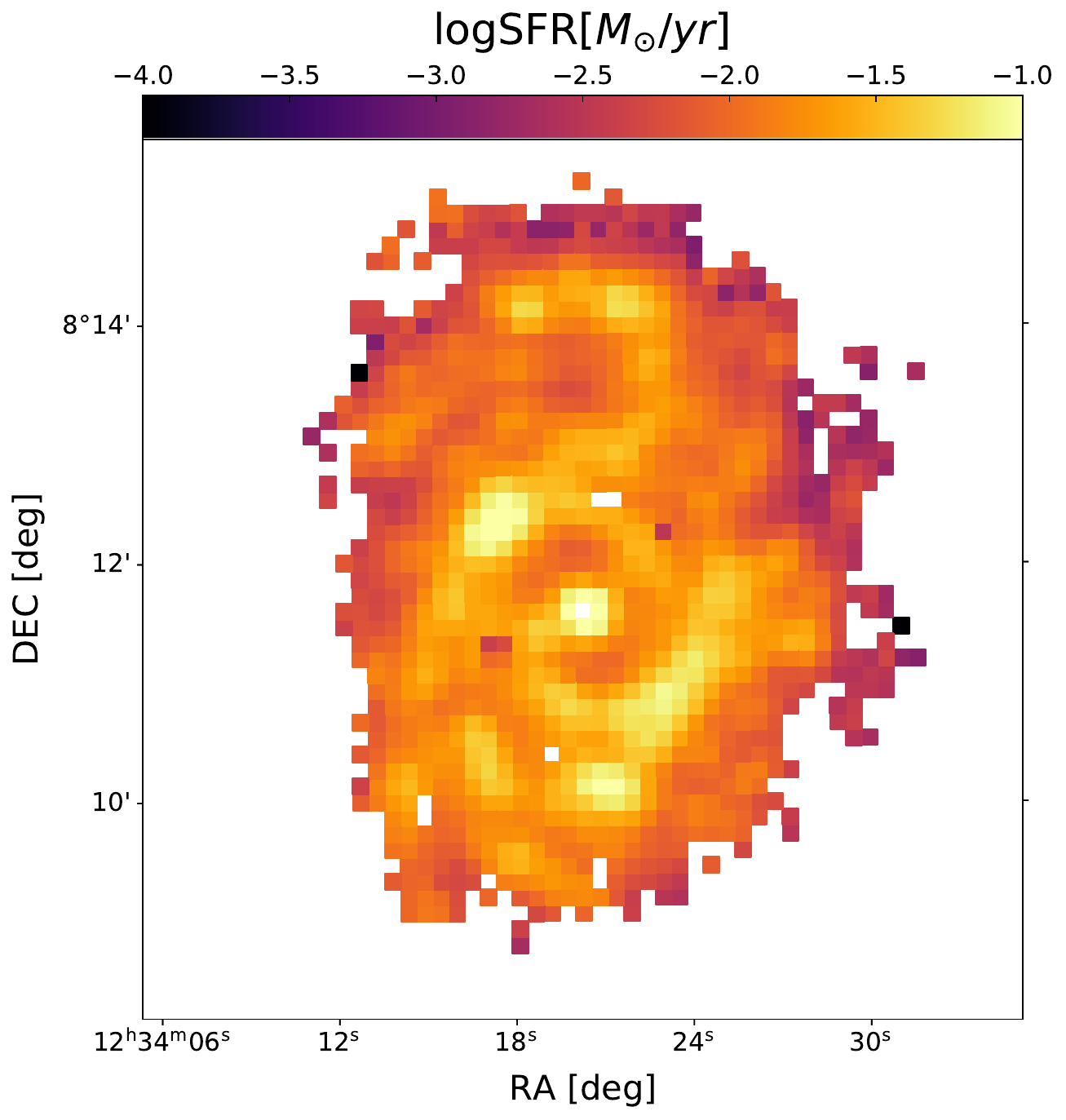}
                
                \label{label2}
        \end{minipage}
\end{figure*}

\begin{figure*}[htpb]
        \centering
        \begin{minipage}{1\columnwidth}
                \centering
                \includegraphics[width=\textwidth]{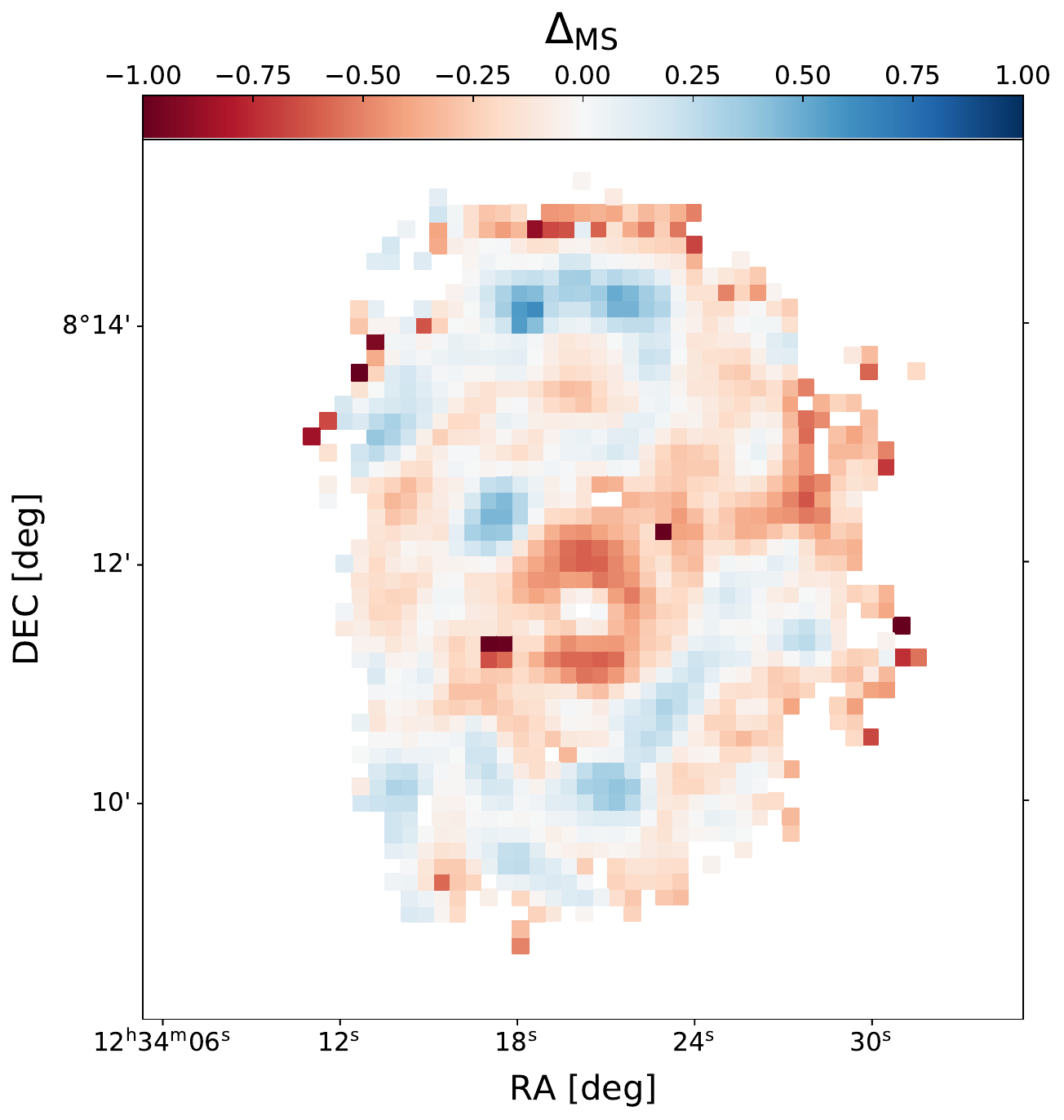}
                
                \label{label3}
        \end{minipage}%
        \begin{minipage}{1\columnwidth}
                \centering
                \includegraphics[width=\textwidth]{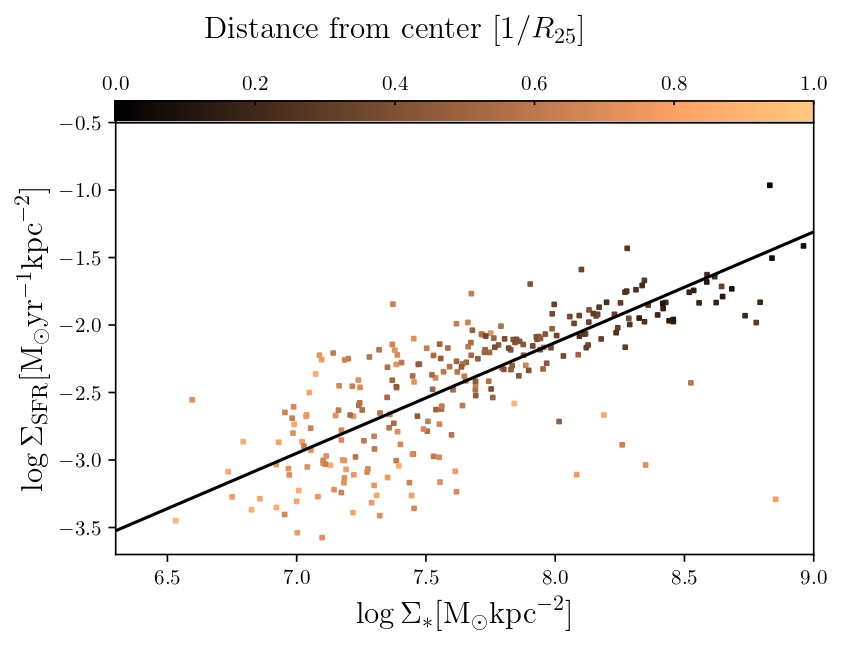}
        
                \label{label4}
        \end{minipage}
        \caption{Same as Fig.\ref{fig:NGC4321} for NGC4535}
\end{figure*}

\begin{figure*}[htpb]
        \centering
        \begin{minipage}{1\columnwidth}
                \centering
                \includegraphics[width=\textwidth]{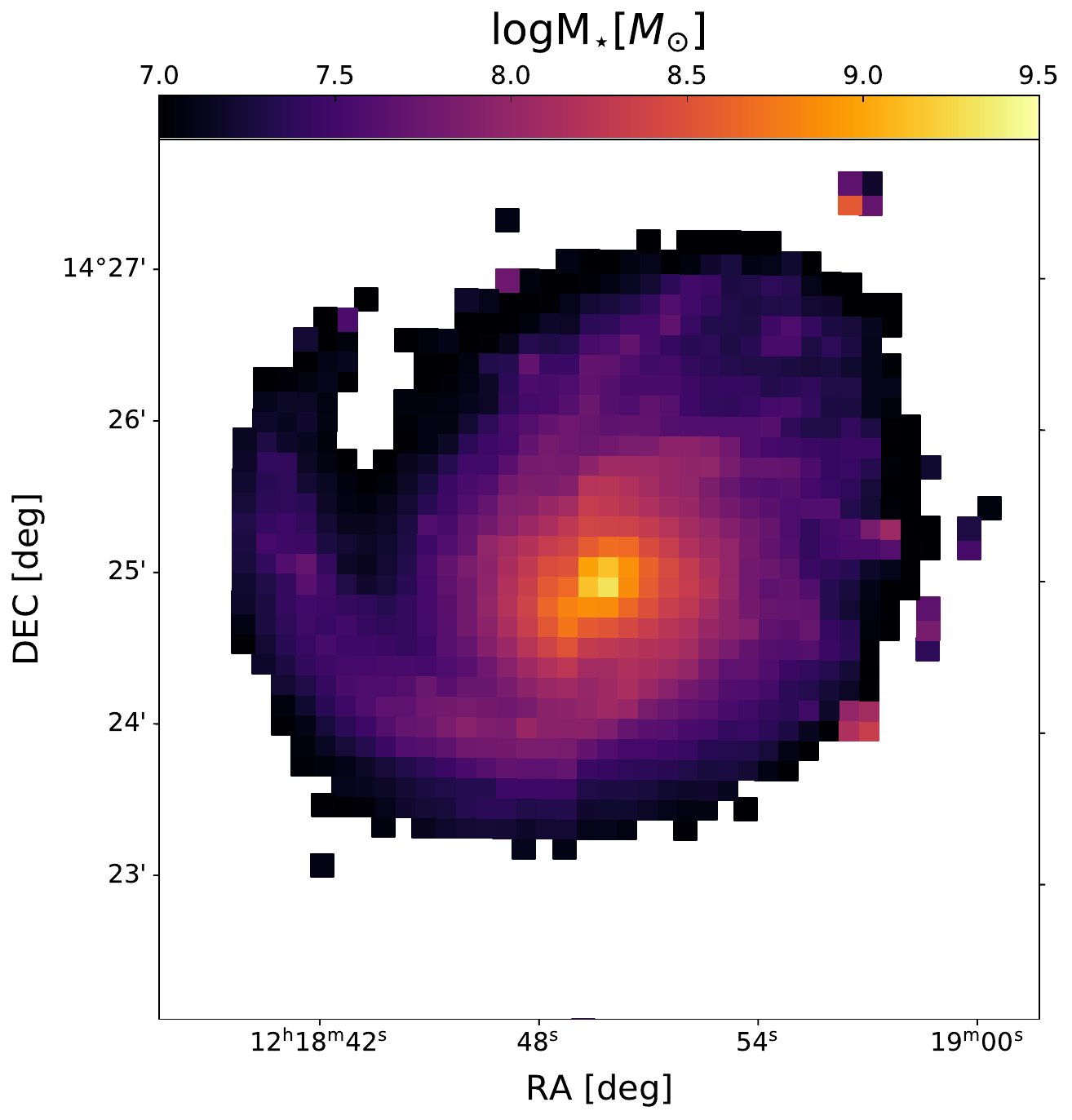}
                
                \label{label1}
        \end{minipage}%
        \begin{minipage}{1\columnwidth}
                \centering
                \includegraphics[width=\textwidth]{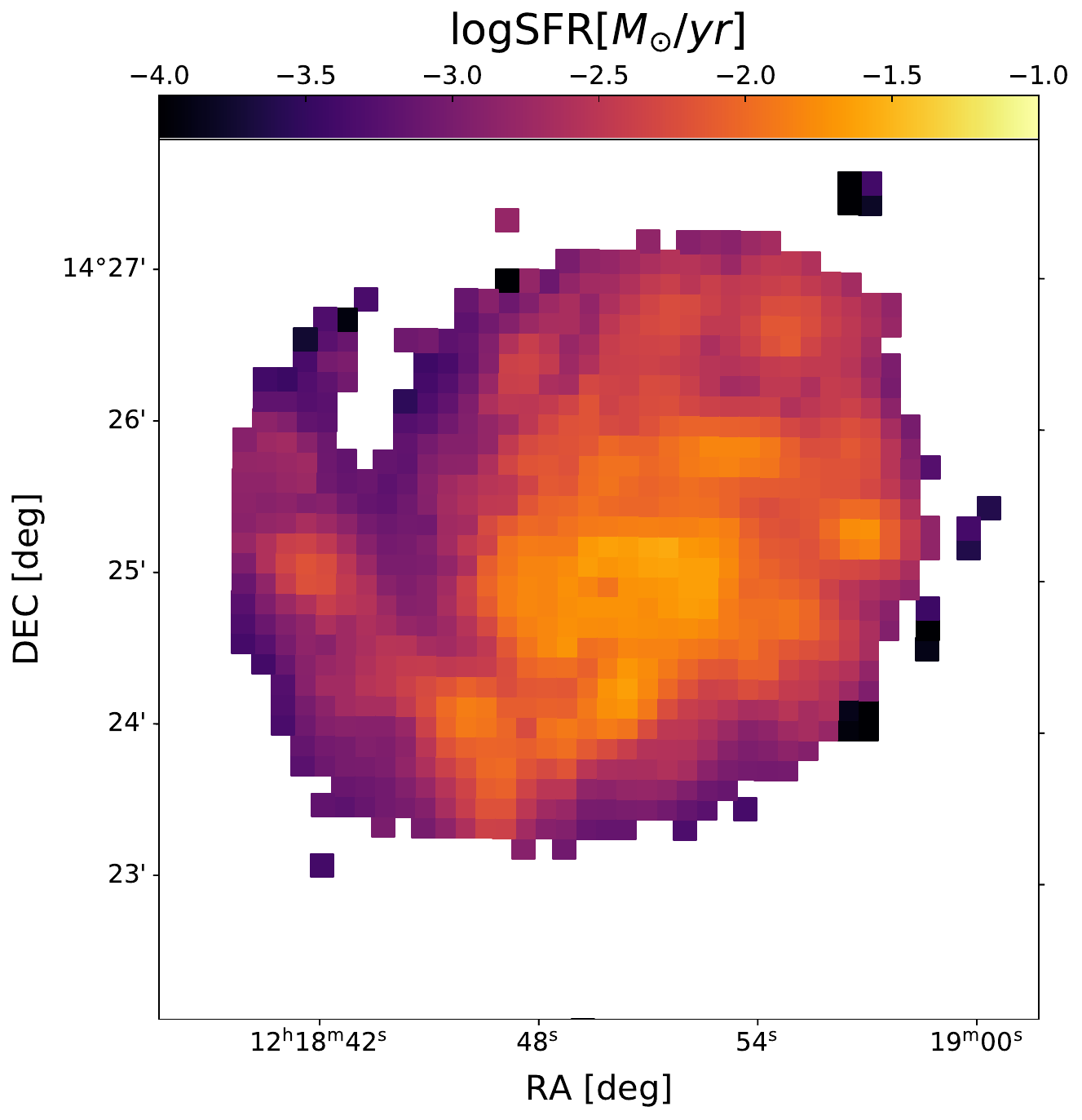}
                
                \label{label2}
        \end{minipage}
\end{figure*}

\begin{figure*}[htpb]
        \centering
        \begin{minipage}{1\columnwidth}
                \centering
                \includegraphics[width=\textwidth]{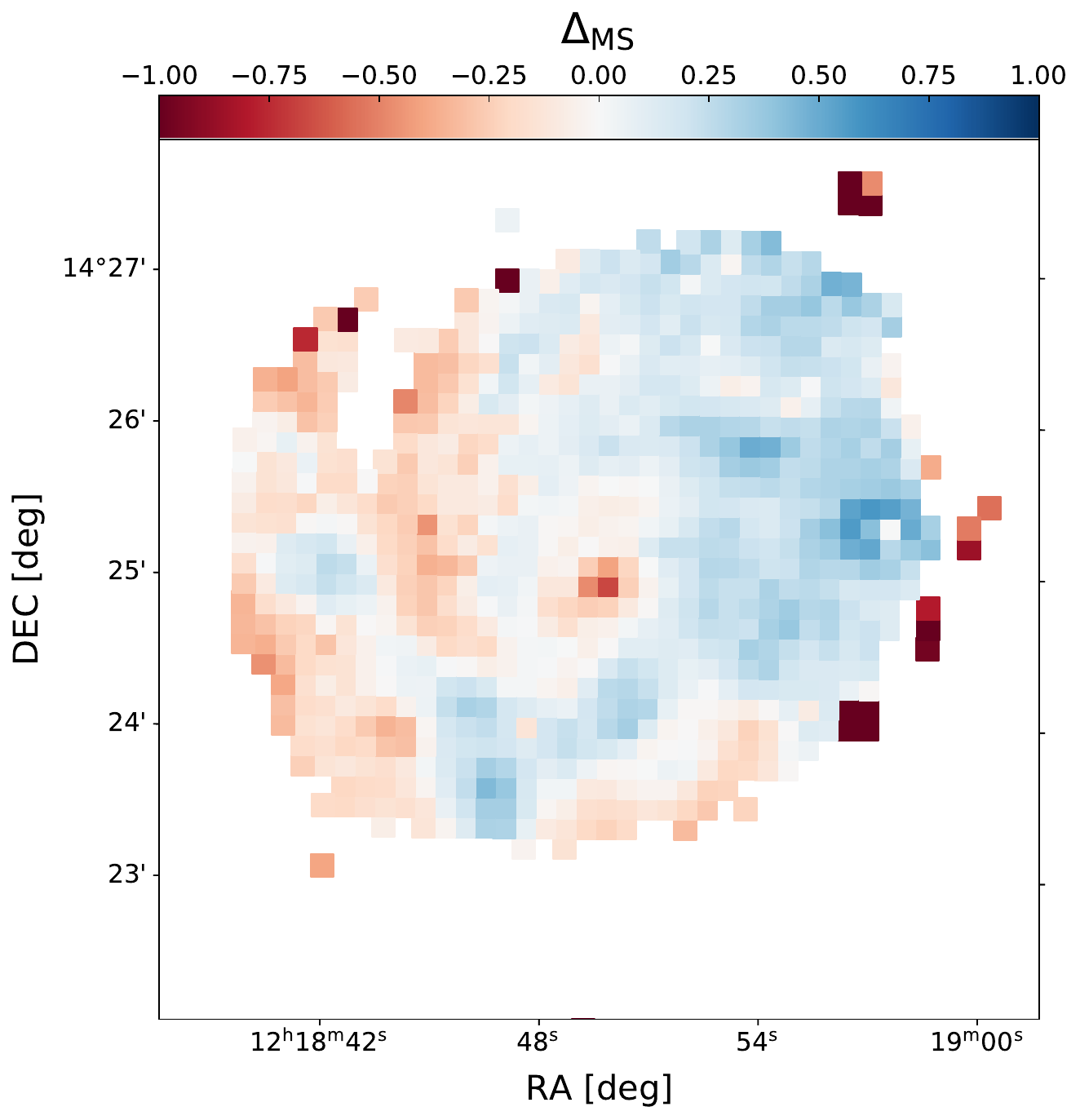}
                
                \label{label3}
        \end{minipage}%
        \begin{minipage}{1\columnwidth}
                \centering
                \includegraphics[width=\textwidth]{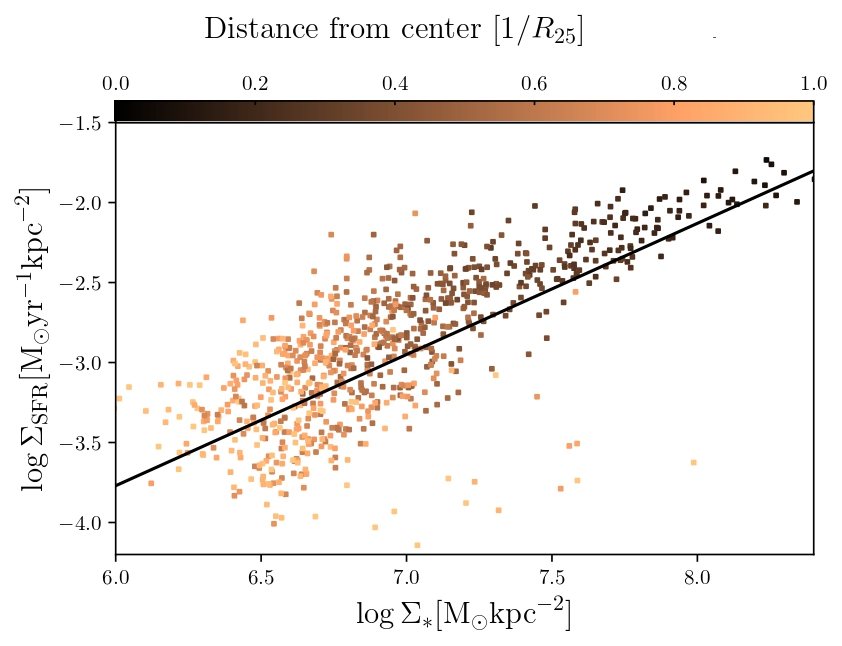}
        
                \label{label4}
        \end{minipage}
        \caption{Same as Fig.\ref{fig:NGC4321} for NGC4254}
\end{figure*}

\begin{figure*}[htpb]
        \centering
        \begin{minipage}{1\columnwidth}
                \centering
                \includegraphics[width=\textwidth]{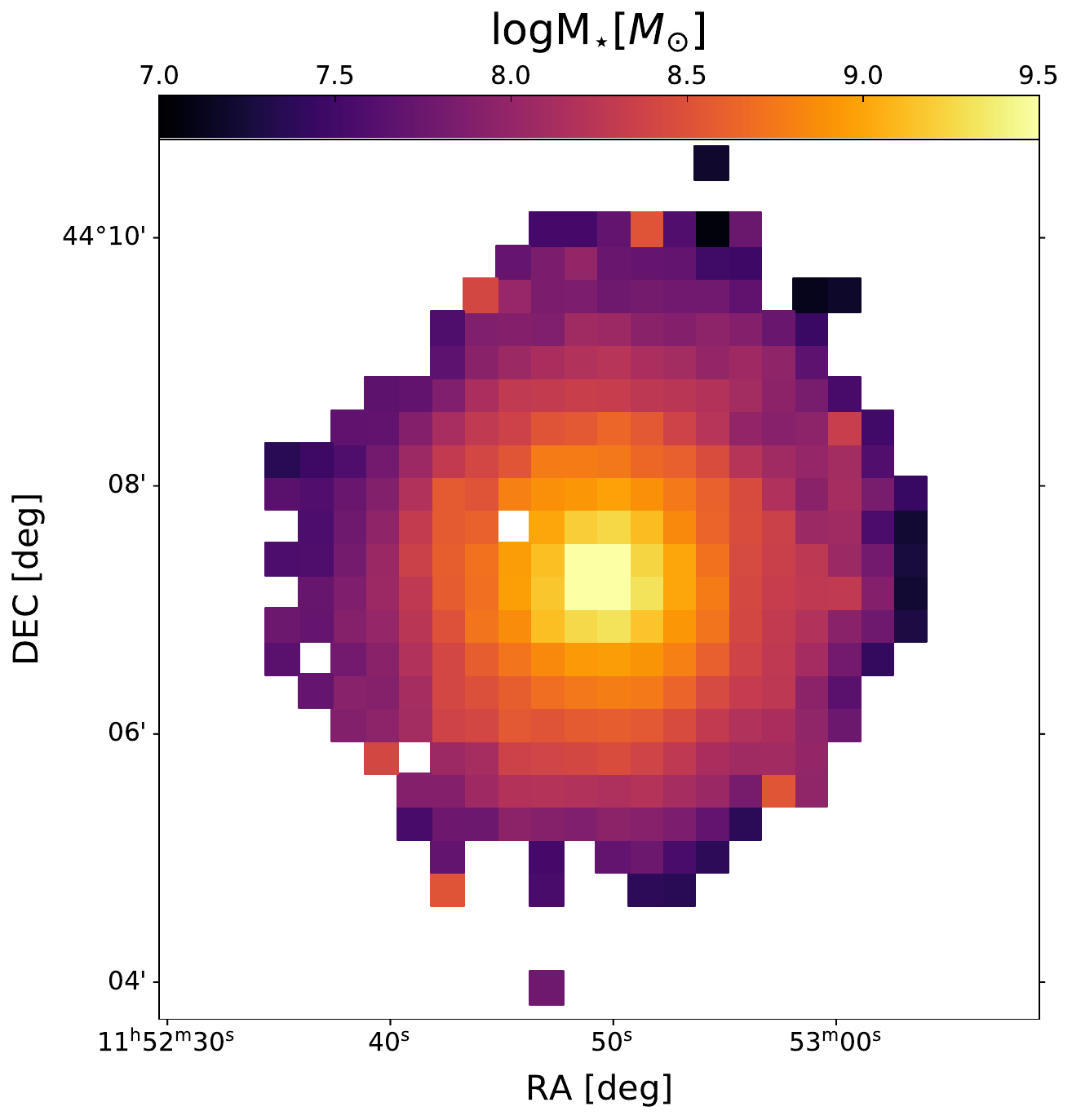}
                
                \label{label1}
        \end{minipage}%
        \begin{minipage}{1\columnwidth}
                \centering
                \includegraphics[width=\textwidth]{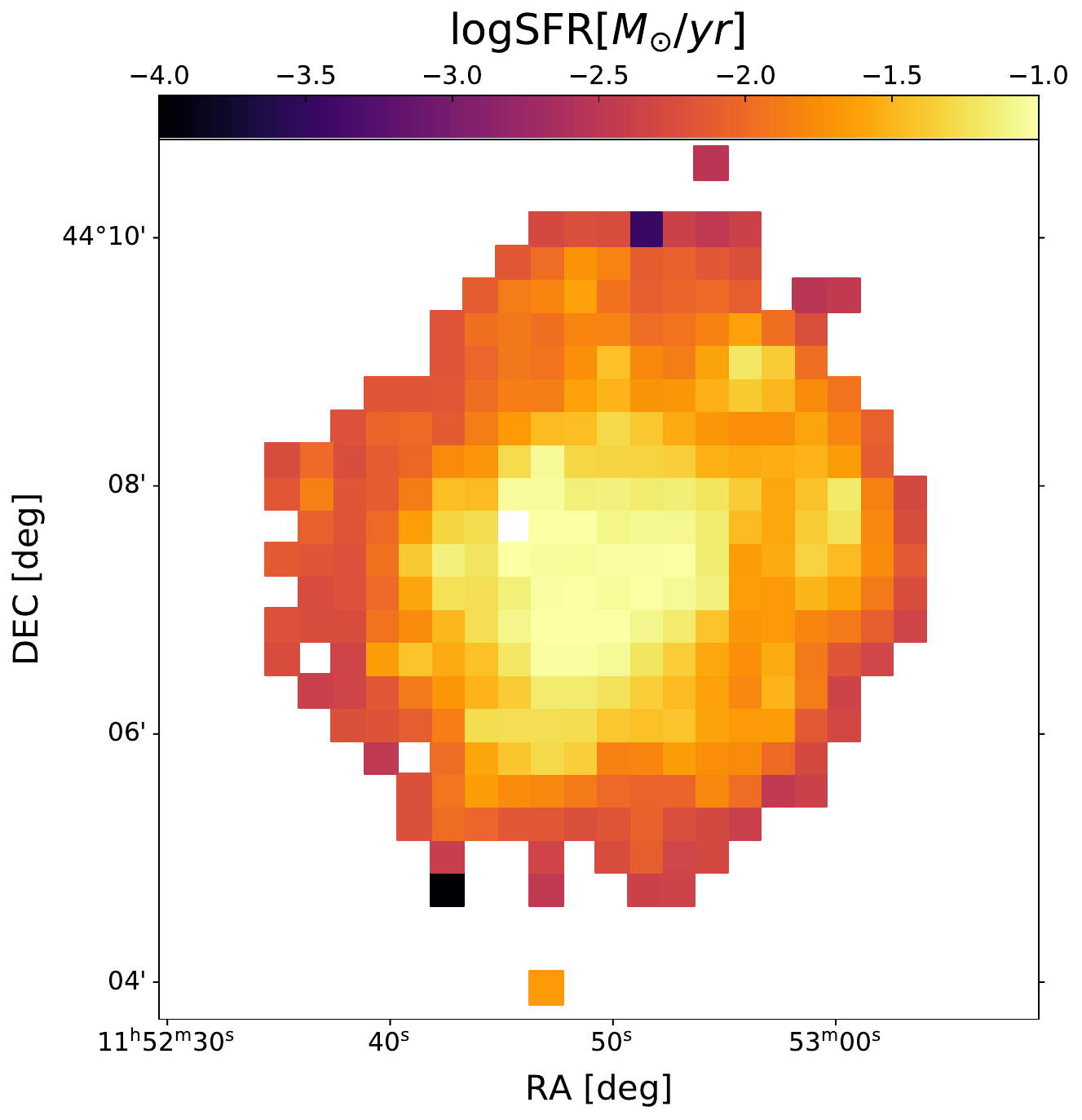}
                
                \label{label2}
        \end{minipage}
\end{figure*}

\begin{figure*}[htpb]
        \centering
        \begin{minipage}{1\columnwidth}
                \centering
                \includegraphics[width=\textwidth]{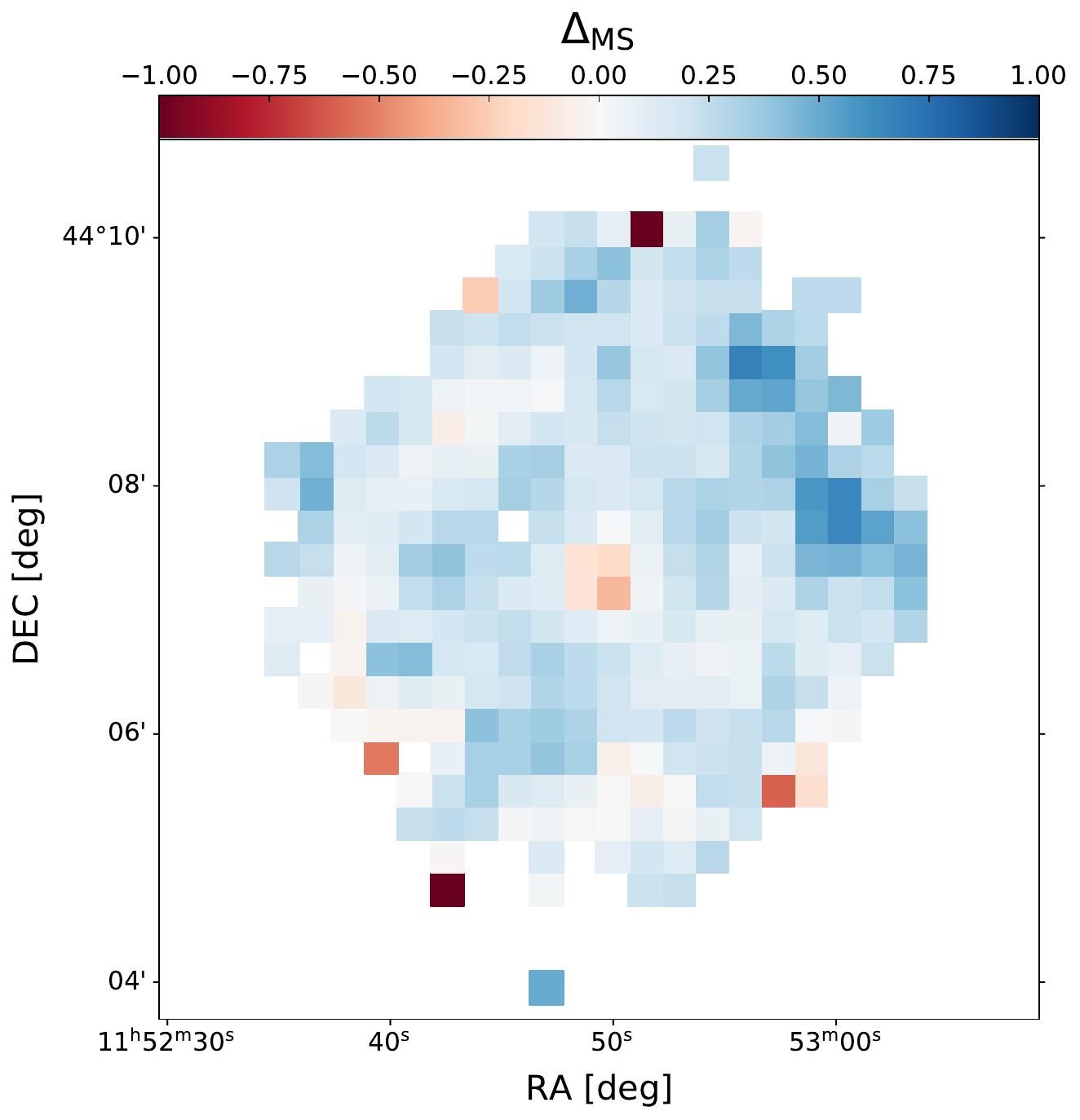}
                
                \label{label3}
        \end{minipage}%
        \begin{minipage}{1\columnwidth}
                \centering
                \includegraphics[width=\textwidth]{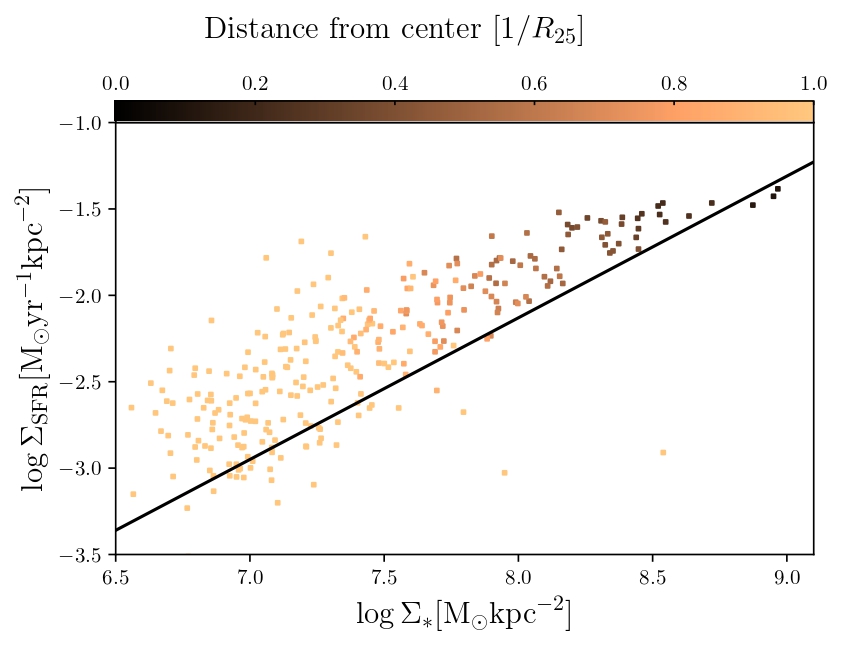}
        
                \label{label4}
        \end{minipage}
        \caption{Same as Fig.\ref{fig:NGC4321} for NGC3938}
\end{figure*}

\begin{figure*}[htpb]
        \centering
        \begin{minipage}{1\columnwidth}
                \centering
                \includegraphics[width=\textwidth]{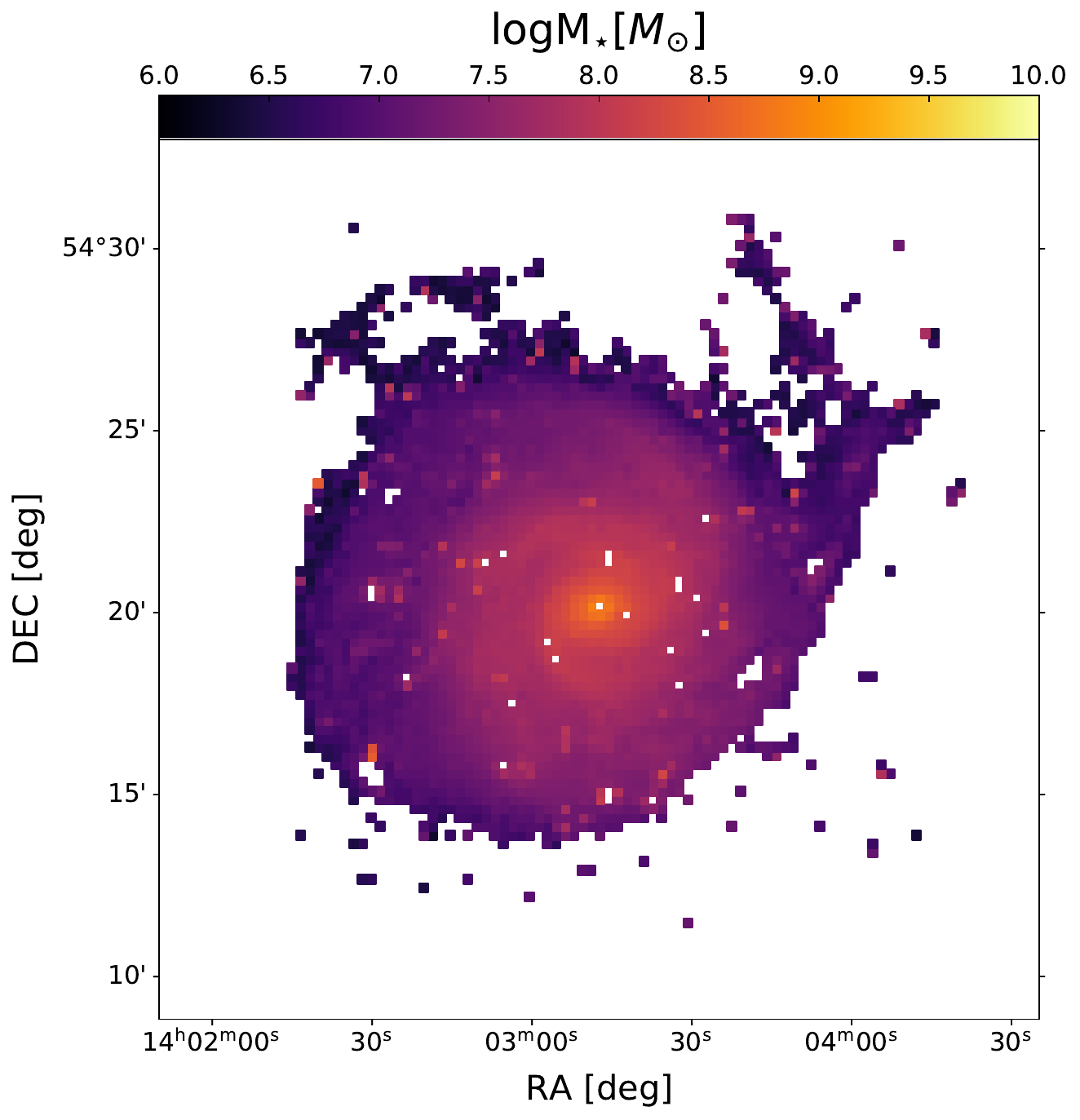}
                
                \label{label1}
        \end{minipage}%
        \begin{minipage}{1\columnwidth}
                \centering
                \includegraphics[width=\textwidth]{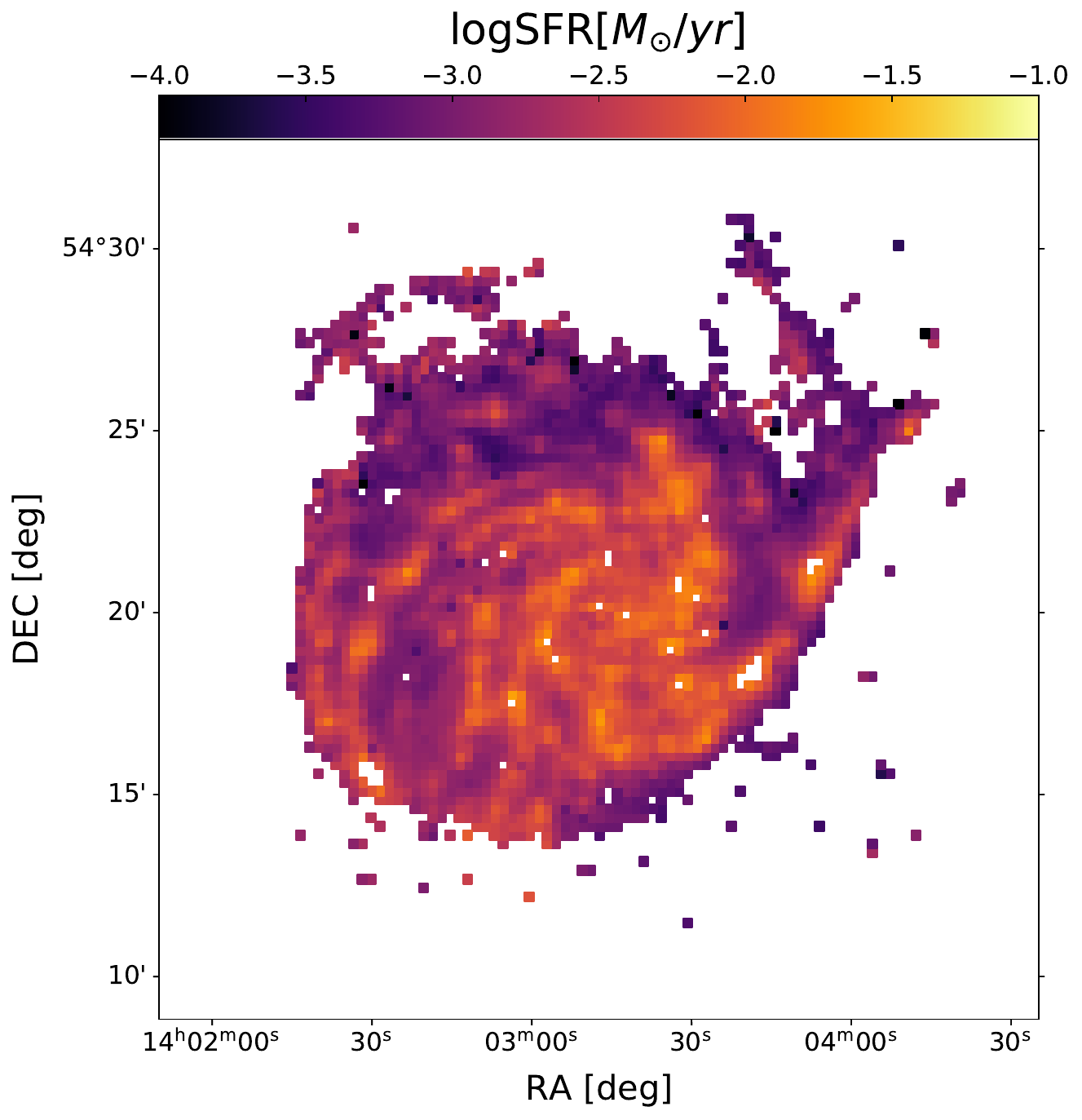}
                
                \label{label2}
        \end{minipage}
\end{figure*}

\begin{figure*}[htpb]
        \centering
        \begin{minipage}{1\columnwidth}
                \centering
                \includegraphics[width=\textwidth]{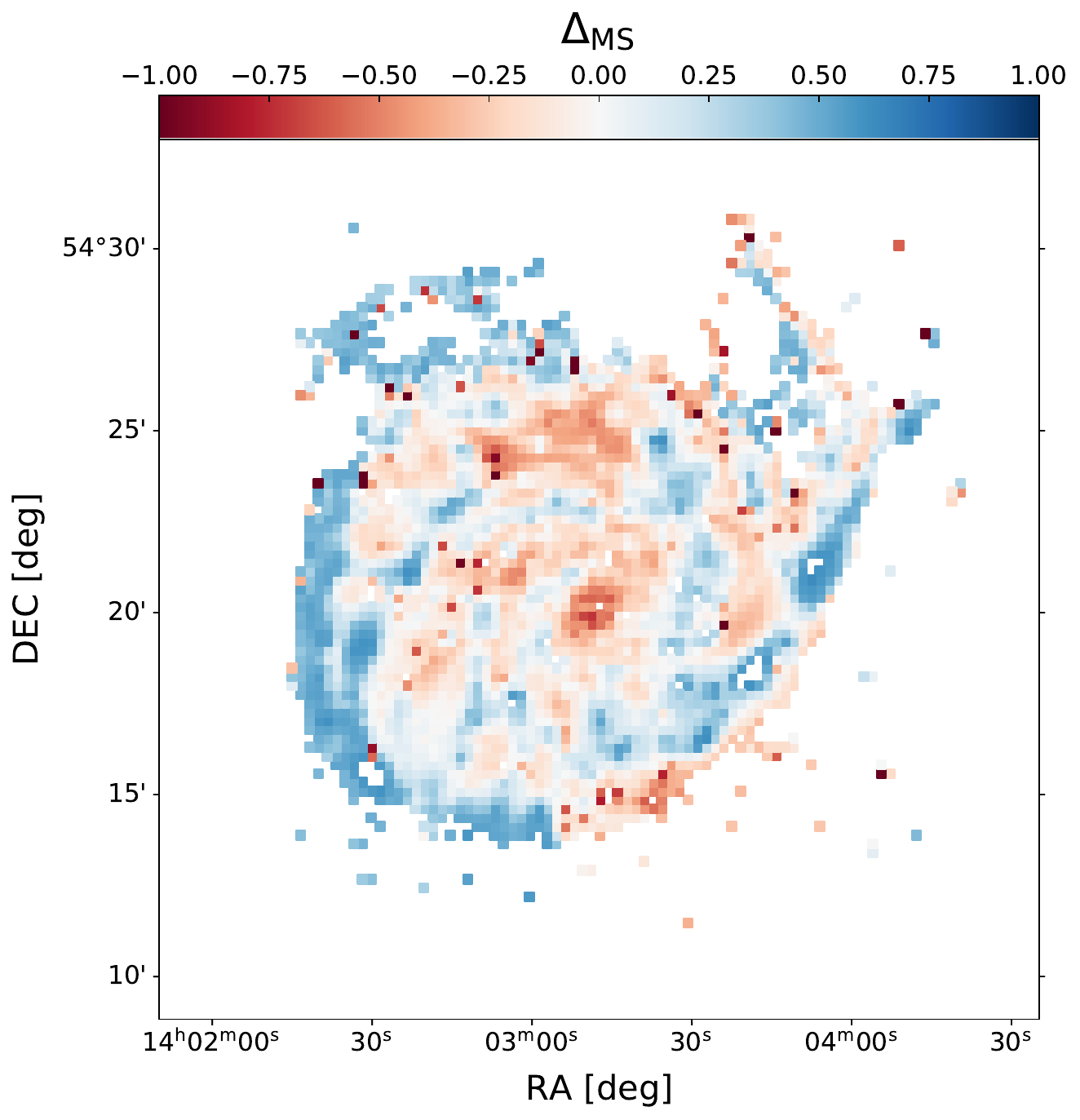}
                
                \label{label3}
        \end{minipage}%
        \begin{minipage}{1\columnwidth}
                \centering
                \includegraphics[width=\textwidth]{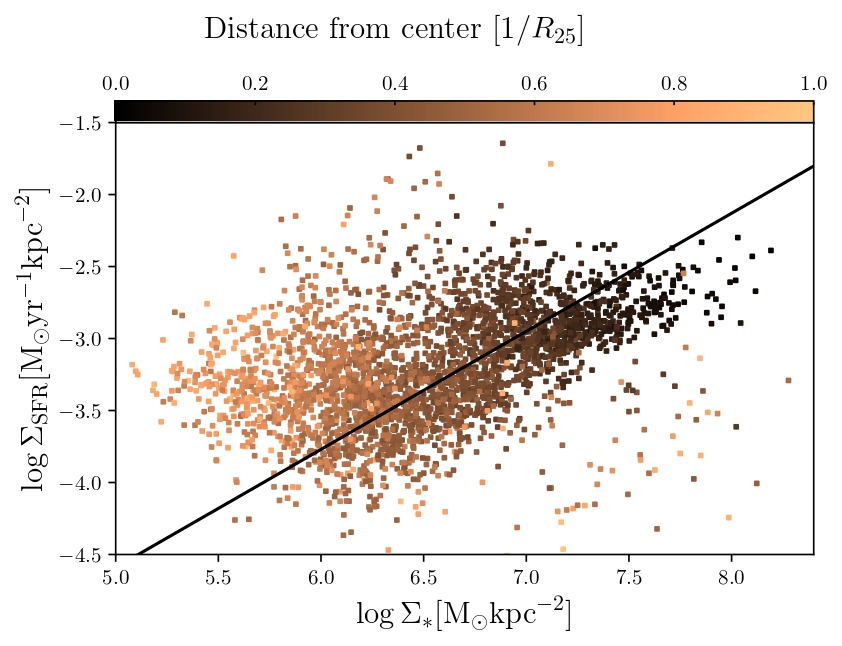}
        
                \label{label4}
        \end{minipage}
        \caption{Same as Fig.\ref{fig:NGC4321} for NGC5457}
\end{figure*}

\newpage

\begin{figure*}[htpb]
        \centering
        \begin{minipage}{1\columnwidth}
                \centering
                \includegraphics[width=\textwidth]{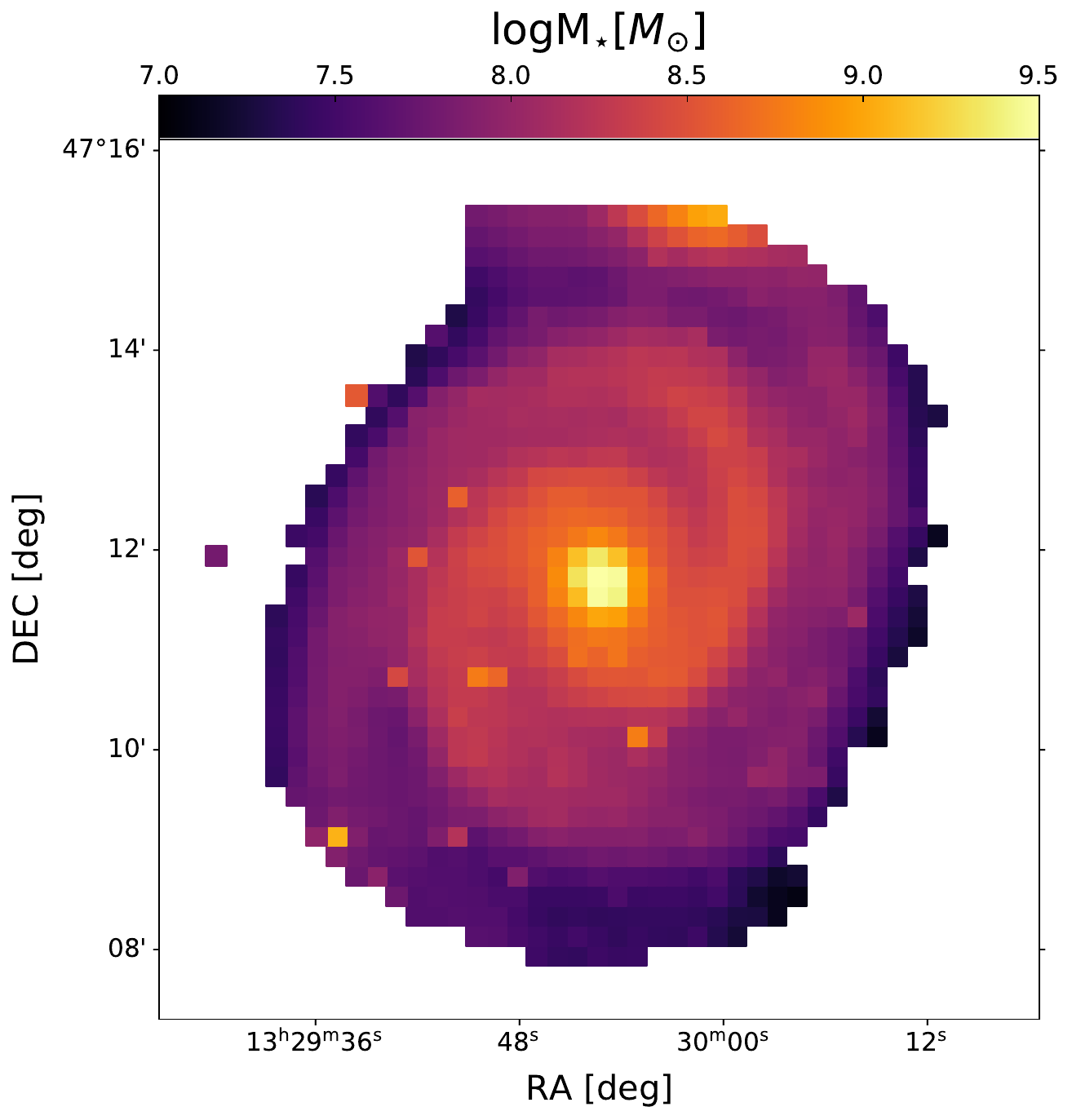}
                
                \label{label1}
        \end{minipage}%
        \begin{minipage}{1\columnwidth}
                \centering
                \includegraphics[width=\textwidth]{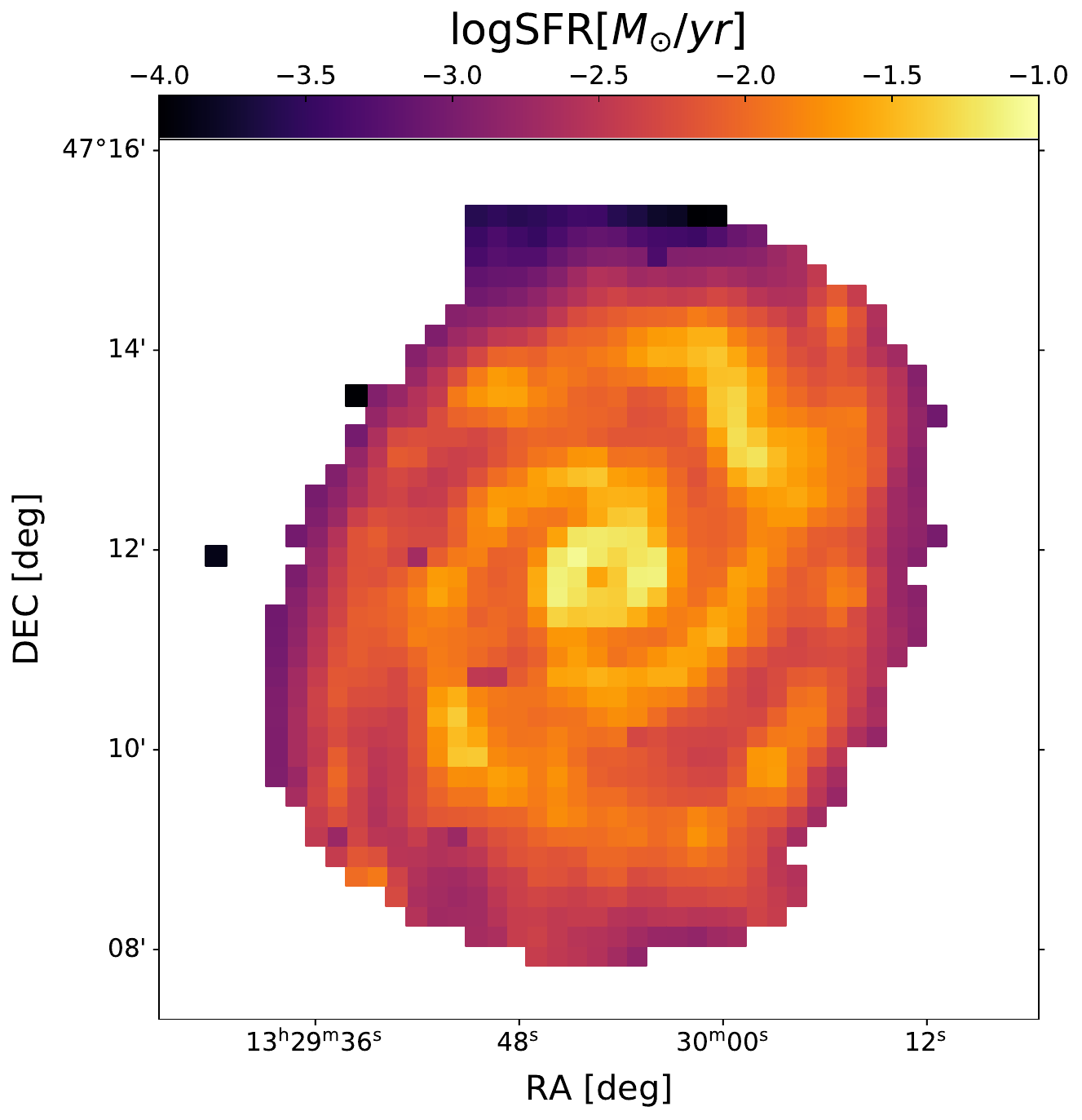}
                
                \label{label2}
        \end{minipage}
\end{figure*}

\begin{figure*}[htpb]
        \centering
        \begin{minipage}{1\columnwidth}
                \centering
                \includegraphics[width=\textwidth]{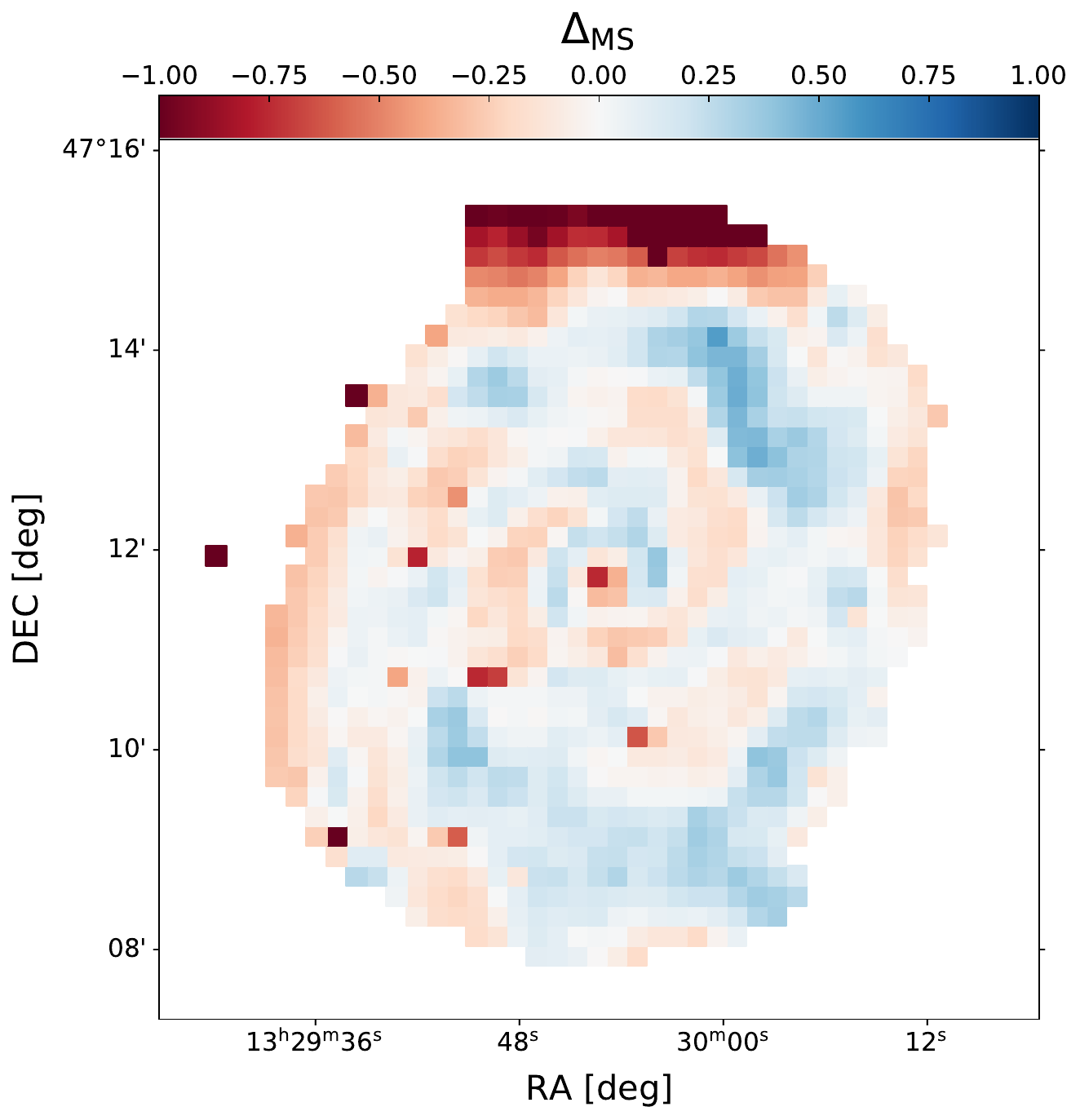}
                
                \label{label3}
        \end{minipage}%
        \begin{minipage}{1\columnwidth}
                \centering
                \includegraphics[width=\textwidth]{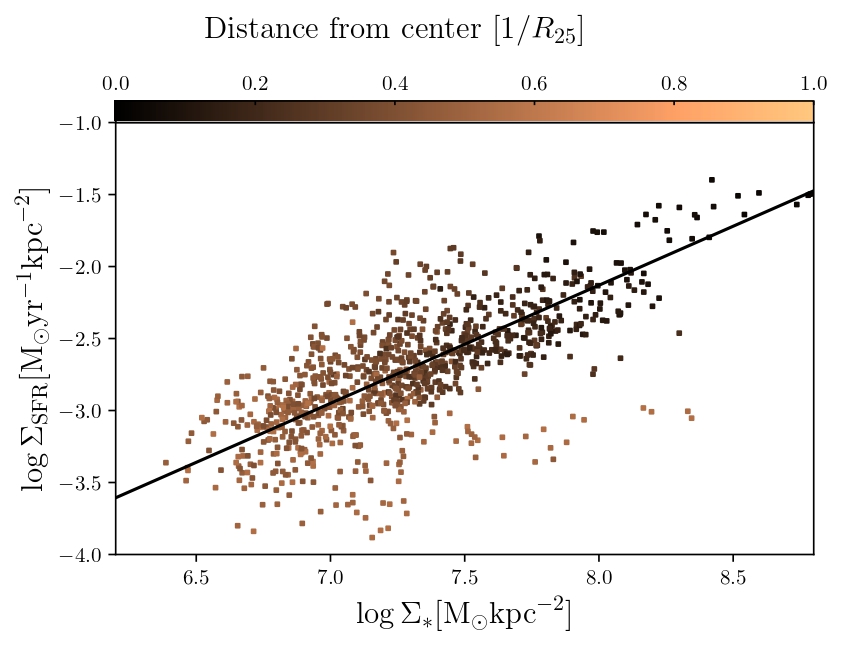}
        
                \label{label4}
        \end{minipage}
        \caption{Same as Fig.\ref{fig:NGC4321} for NGC5194}
\end{figure*}

\newpage

\begin{figure*}[htpb]
        \centering
        \begin{minipage}{1\columnwidth}
                \centering
                \includegraphics[width=\textwidth]{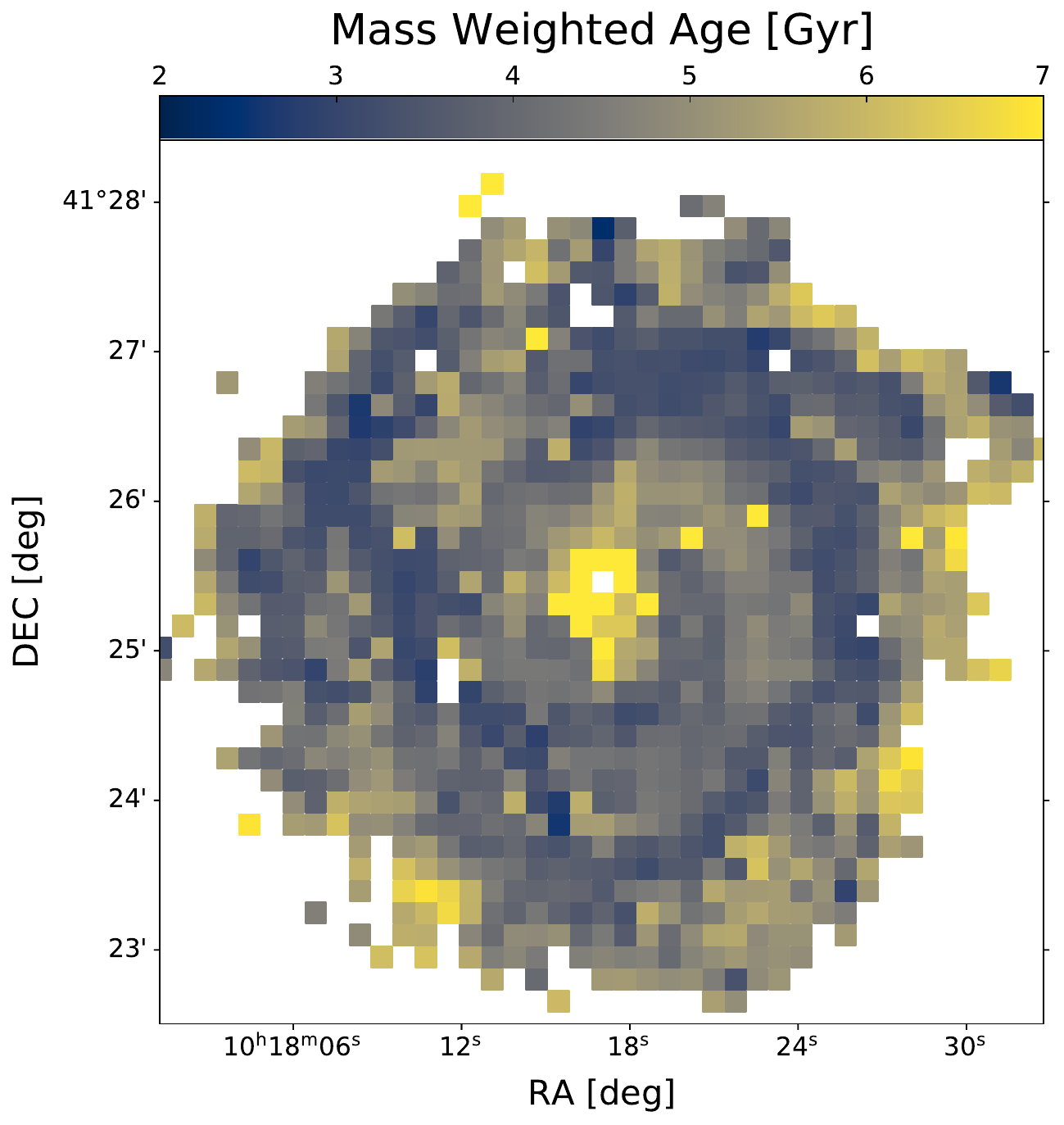}
                
                \label{label1}
        \end{minipage}%
        \begin{minipage}{1\columnwidth}
                \centering
                \includegraphics[width=\textwidth]{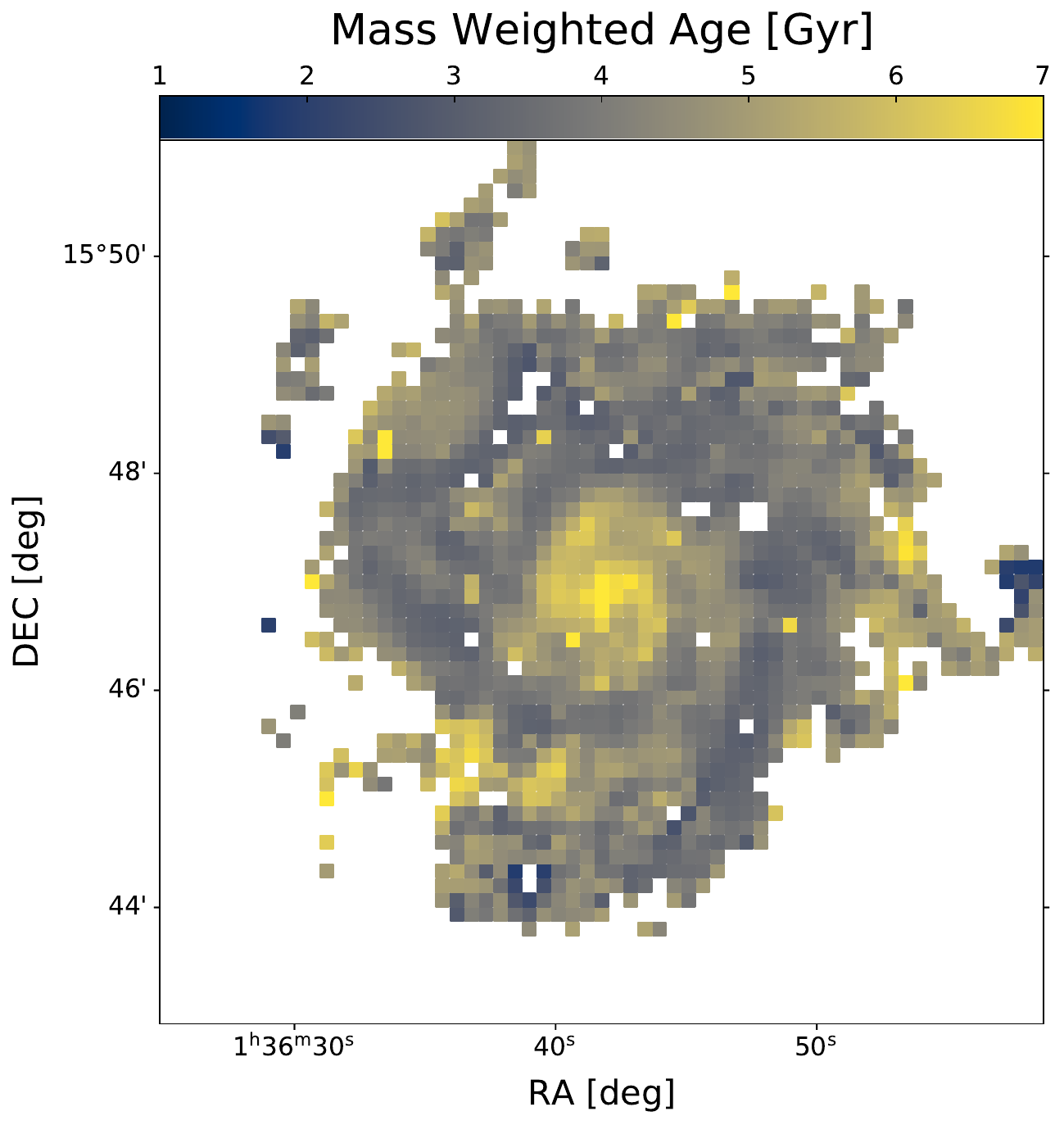}
                
                \label{label2}
        \end{minipage}
        \caption{Mass-weighted age for NGC3184 and NGC0628}
\end{figure*}

\begin{figure*}[htpb]
        \centering
        \begin{minipage}{1\columnwidth}
                \centering
                \includegraphics[width=\textwidth]{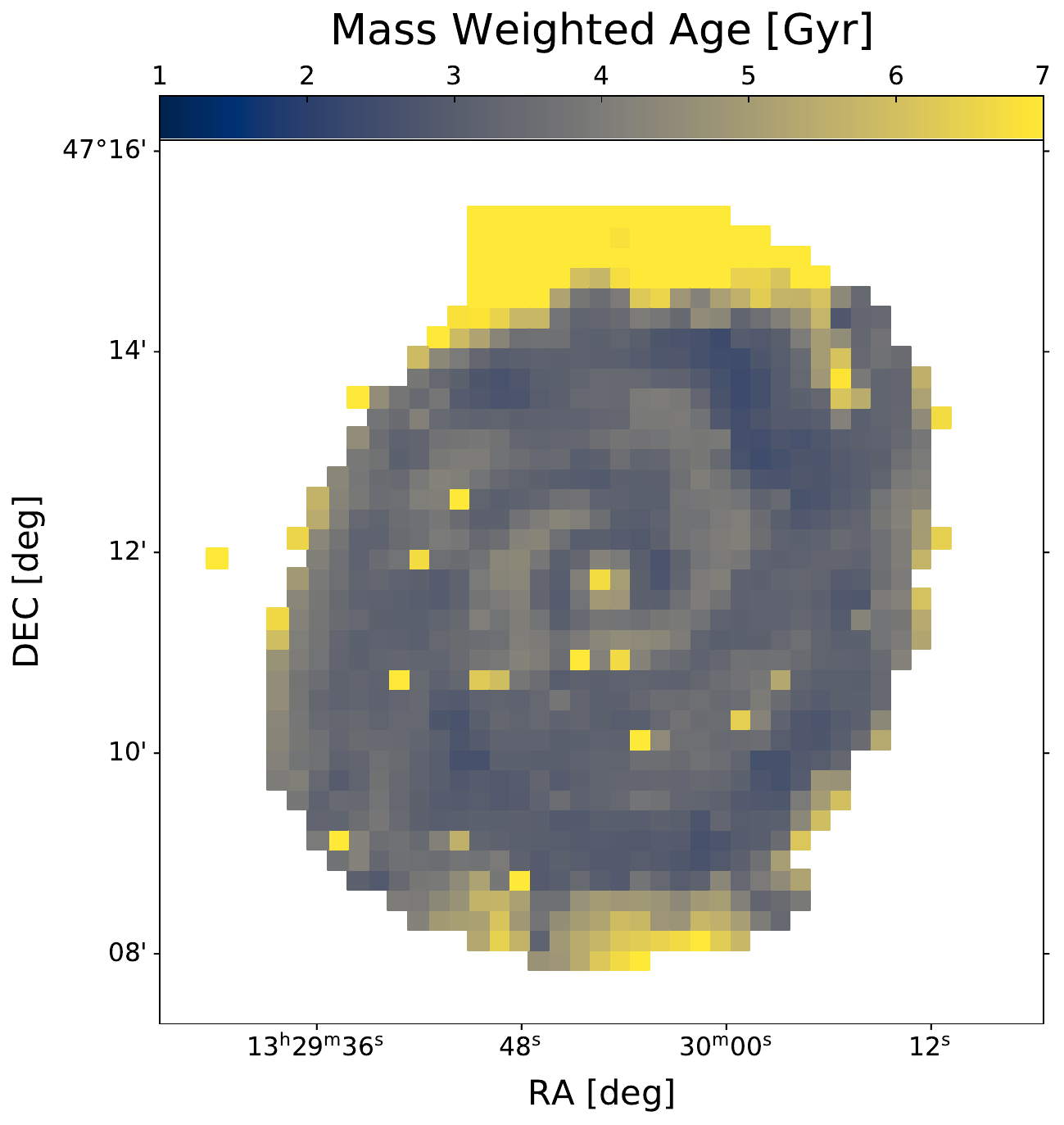}
                
                \label{label1}
        \end{minipage}%
        \begin{minipage}{1\columnwidth}
                \centering
                \includegraphics[width=\textwidth]{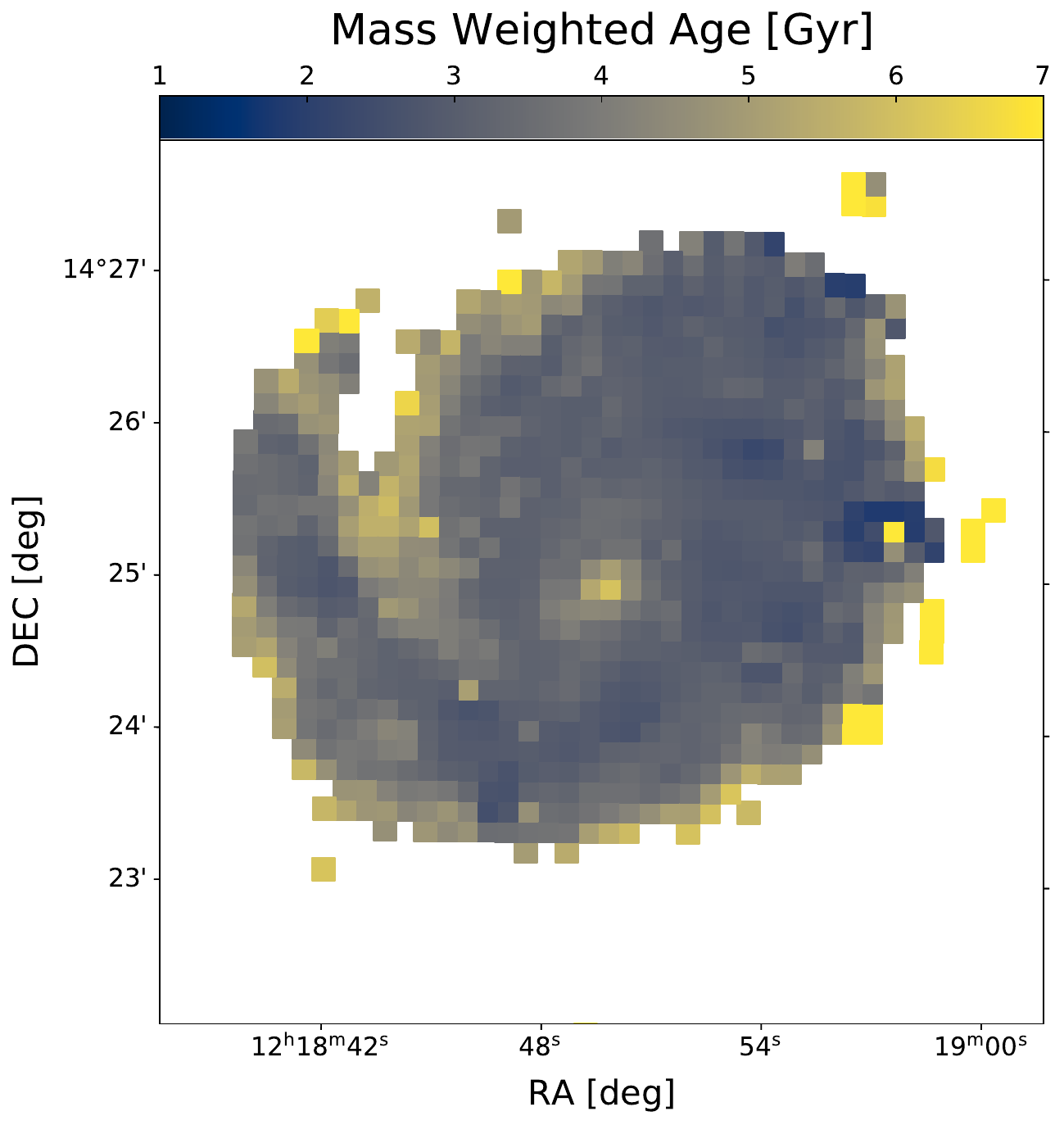}
                
                \label{label2}
        \end{minipage}
        \caption{Mass-weighted age for NGC5194 and NGC4254}
\end{figure*}

\newpage

\begin{figure*}[htpb]
        \centering
        \begin{minipage}{1\columnwidth}
                \centering
                \includegraphics[width=\textwidth]{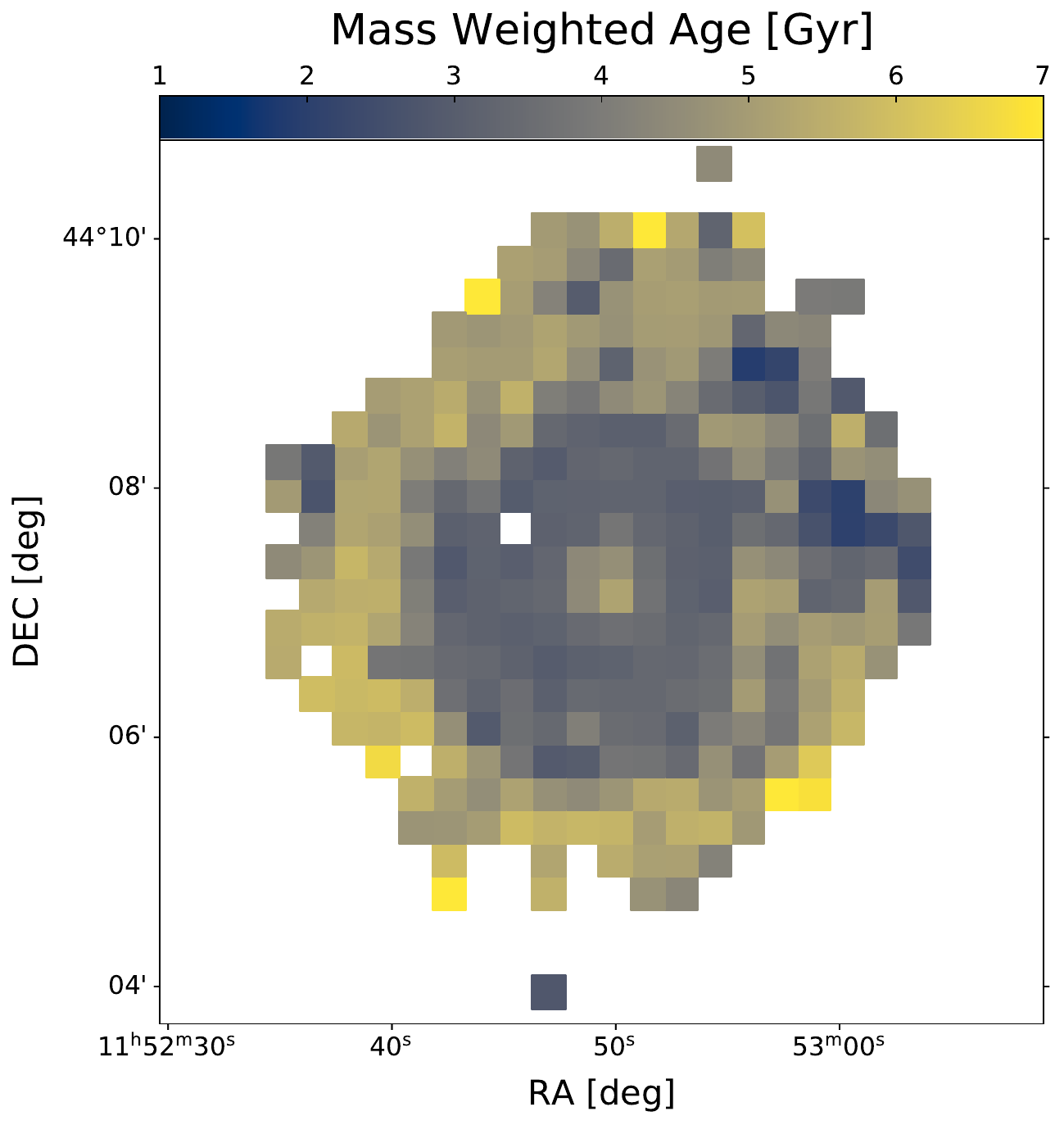}
                
                \label{label1}
        \end{minipage}%
        \begin{minipage}{1\columnwidth}
                \centering
                \includegraphics[width=\textwidth]{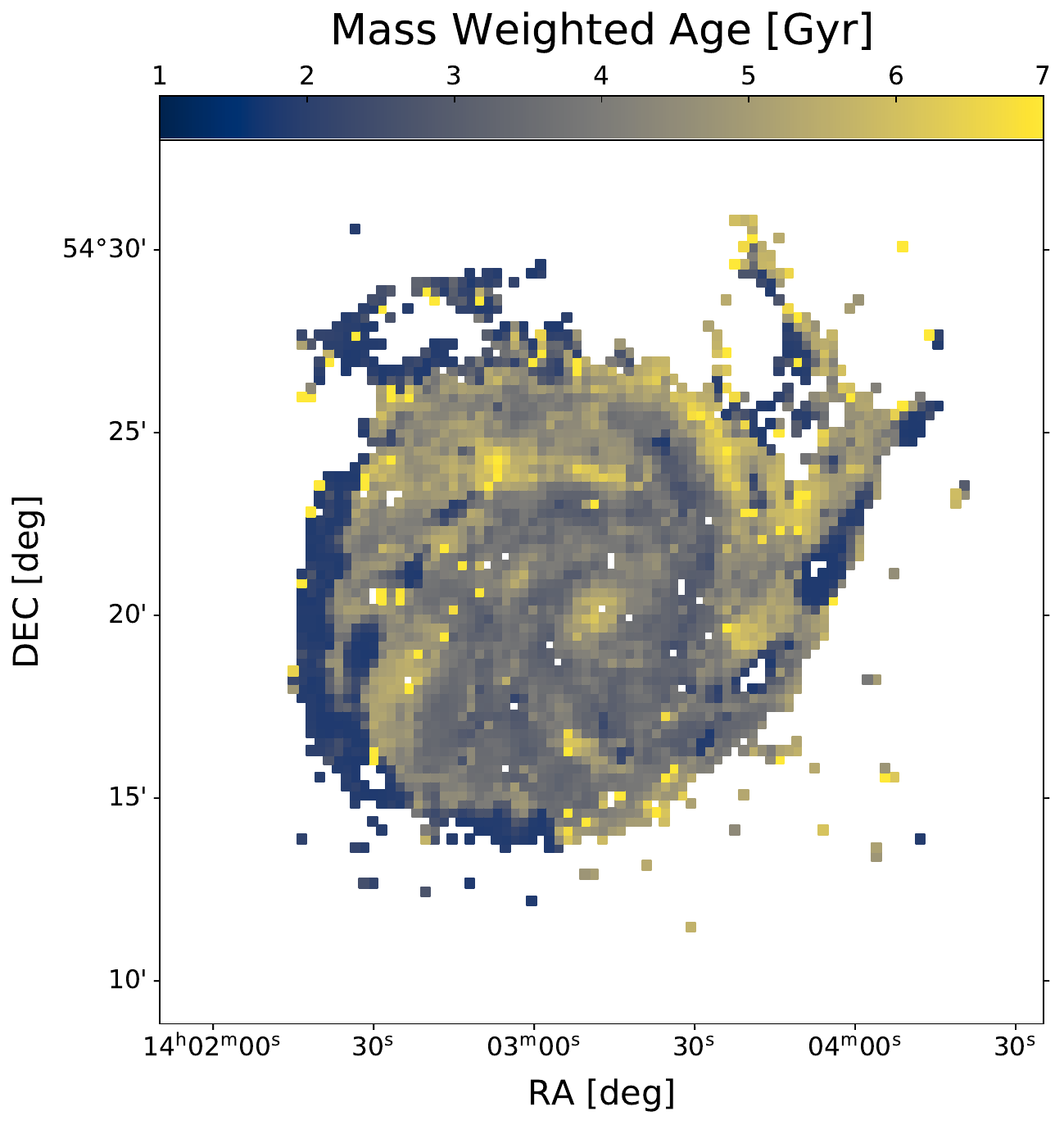}
                
                \label{label2}
        \end{minipage}
        \caption{Mass-weighted age for NGC3938 and NGC5457}
\end{figure*}

\begin{figure*}[htpb]
        \centering
        \begin{minipage}{1\columnwidth}
                \includegraphics[width=\textwidth]{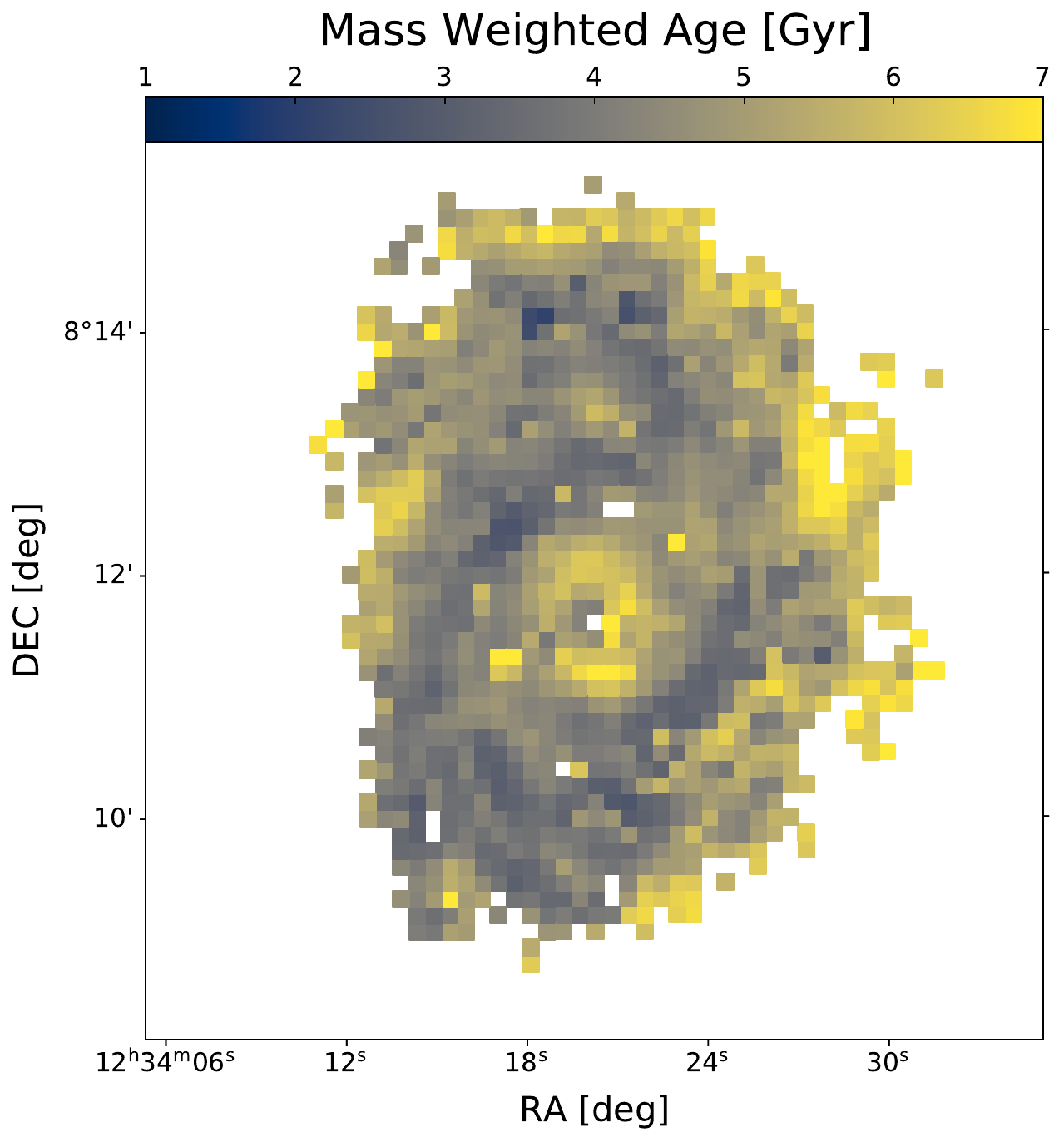}
                
                \label{label1}
        \end{minipage}%
        \caption{Mass-weighted age for NGC4535}
\end{figure*}

\end{document}